\documentclass[10pt,a4paper,twoside,fleqn]{ActaStyle}
\usepackage{graphicx}
\usepackage{times}
\usepackage{cite}
\usepackage{calrsfs}

\def\authorheading{Z. Stuchl\'{\i}k, S. Hled\'{\i}k}
\def\shorttitle{Reissner--Nordstr\"om spacetimes
  with a nonzero cosmological constant}

\newcommand{\cc}{cosmological constant}
\newcommand{\ed}{embedding diagrams}
\newcommand{\nnd}{\discretionary{--}{--}{--}}
\newcommand{\Schw}{Schwarz\-schild}
\newcommand{\RN}{Reiss\-ner\discretionary{--}{--}{--}Nordstr\"om}
\newcommand{\dS}{de~Sitter}
\newcommand{\adS}{anti\discretionary{-}{-}{-}de~Sitter}
\newcommand{\aodS}{(anti\discretionary{-}{-}{-})de~Sitter}
\newcommand{\beq}{\begin{equation}} \newcommand{\eeq}{\end{equation}}
\newcommand{\bea}{\begin{eqnarray}} \newcommand{\eea}{\end{eqnarray}}
\newcommand{\diff}{\text{d}}
\newcommand{\Diff}{\text{D}}
\newcommand{\e}{\text{e}}

\newcommand{\four}{four\discretionary{-}{-}{-}}
\newcommand{\tfrac}[2]{\textstyle\frac{#1}{#2}}
\newcommand{\p}{\partial}
\newcommand{\oder}[2]{\frac{\diff #1}{\diff #2}}
\newcommand{\pder}[2]{\frac{\partial #1}{\partial #2}}
\makeatletter
\def\sign{\mathop{\operator@font sign}\nolimits}
\makeatother
\def\vec#1{%
  \ifmmode%
    \mathchoice{\mbox{\boldmath$\displaystyle#1$}}
    {\mbox{\boldmath$\textstyle#1$}}
    {\mbox{\boldmath$\scriptstyle#1$}}
    {\mbox{\boldmath$\scriptscriptstyle#1$}}%
  \else
    \hbox{\boldmath$\textstyle#1$}%
  \fi}
\newcommand{\Md}{---}
\let\text=\mathrm
\begin{document}
\pagerange{1}{45}
\title{PROPERTIES OF THE REISSNER--NORDSTR\"OM SPACETIMES\\
  WITH A NONZERO COSMOLOGICAL CONSTANT%
  \footnote{Published in Acta Physica Slovaca \textbf{52}(5) (2002),
    pp.~363--407.}}
\author{Z. Stuchl\'{\i}k\email{zdenek.stuchlik@fpf.slu.cz},
        S. Hled\'{\i}k\email{stanislav.hledik@fpf.slu.cz}}%
{%
  Institute of Physics,
  Faculty of Philosophy and Science,
  Silesian University in Opava,
  Bezru\v{c}ovo n\'am.~13,
  CZ-746\,01~Opava, Czech Republic%
}
\day{June 12, 2002}

\abstract{%
  Properties of the \RN{} black-hole and naked-singularity spacetimes with
  a nonzero cosmological constant $\Lambda$ are represented by their
  geodetical structure and embedding diagrams of the central planes of both
  the ordinary geometry and associated optical reference geometry. Both the
  asymptotically \dS{} ($\Lambda > 0$) and \adS{} ($\Lambda <0$) spacetimes
  are considered and compared with the asymptotically flat ($\Lambda = 0$)
  spacetimes.  Motion of test particles (timelike geodesics for uncharged
  particles) and photons (null geodesics) is described in terms of an
  appropriate `effective potential.' Circular geodesics are discussed and
  photon escape cones are determined. The spacetimes are divided in their
  parameter space into separated parts according to different character of
  the effective potential and properties of the circular geodesics. In all
  asymptotically \adS{} black-hole spacetimes and some asymptotically \dS{}
  black-hole spacetimes a region containing stable circular geodesics
  exists, which allows accretion processes in the disk regime. On the other
  hand, around some naked singularities, both asymptotically \dS{} and
  \adS{}, even two separated regions with stable circular geodesics exist.
  The inner region is limited from below by particles with zero angular
  momentum that are located in stable equilibrium positions. The inertial
  and gravitational forces related to the optical reference geometry are
  introduced and specified for the circular motion.  It is shown how the
  properties of the centrifugal force are closely related to the properties
  of the optical reference geometry embedding diagrams.}


\section{Introduction}

Recently acquired data from a wide range of cosmological tests, including
the measurements of the present value of the Hubble parameter and dynamical
estimates of the present energy density of the Universe, measurements of
the anisotropy of the cosmic relict radiation, statistics of gravitational
lensing of quasars and active galactic nuclei, galaxy number counts, and
measurements of high-redshift supernovae, indicate convincingly that in the
framework of the inflationary paradigm, a very small relict repulsive
cosmological constant $\Lambda > 0$, or an analogous concept of
quintessence, has to be invoked in order to explain the dynamics of the
recent Universe
\cite{Kra-Tur:1995:GENRG2:,%
      Ost-Ste:1995:NATURE:,%
      Bah-etal:1999:SCIEN:,%
      Cal-Dav-Ste:1998:PHYRL:,%
      Wan-etal:2000:ASTRJ2:,%
      ArP-Muk-Ste:2000:PHYRL:}.
The presence of a repulsive cosmological constant changes dramatically
the asymptotic structure of black-hole backgrounds. Such backgrounds
become asymptotically \dS{} spacetimes, not flat spacetimes.

On the other hand, the \adS{} spacetimes play a crucial role in the
framework of superstring theories
\cite{Sen:1999:29HEP:,Schm:2000:hep-th9912156:}.  Therefore, it is also
important to obtain information on the influence of an attractive \cc{}
($\Lambda < 0$) on the structure of black-hole backgrounds.

It is crucial to understand the role of a nonzero \cc{} in astrophysically
relevant situations. For these purposes, analysis of geodetical motion is
among the most fundamental techniques. Further, the curvature of the
spacetime under consideration can be conveniently demonstrated by using
embedding diagrams of 2-dimensional, appropriately chosen, spacelike
surfaces into 3-dimensional Euclidean space \cite{Mis-Tho-Whe:1973:Gra:}.
The 3-dimensional optical reference geometry
\cite{Abr-Car-Las:1988:GENRG2:} associated with the geometry of the
spacetime under consideration enables a natural `Newtonian' concept of
gravitational and inertial forces and reflects some hidden properties of
the geodetical motion \cite{Abr:1992:MONNR:}. Usually, the \ed{} of
appropriately chosen 2-dimensional sections of the optical reference
geometry directly illustrate properties of the centrifugal force
\cite{Kri-Son-Abr:1998:GENRG2:,%
      Stu-Hle:1999:CLAQG:,%
      Stu-Hle:1999:PHYSR4:,%
      Stu-Hle-Jur:2000:CLAQG:,%
      Stu-etal:2001:PHYSR4:}.
    
{\sloppy
Both the geodetical structure and \ed{} were studied extensively in the
\Schw\nnd\dS{} and \Schw\nnd\adS{} spacetimes \cite{Stu-Hle:1999:PHYSR4:}.
However, all of these spacetimes containing a static region are black-hole
spacetimes. Clearly, it is useful to have an idea of the influence of a
nonzero \cc{} on the character of naked-singularity spacetimes, too.
\par}

Naked singularities, i.e., spacetime singularities not hidden behind an
event horizon, represent solutions to the Einstein equations for the most
exotic compact gravitational objects that could be conceivable to explain
quasars and active galactic nuclei. The Penrose conjecture of cosmic
censorship \cite{Pen:1969:NUOC2:} suggests that no naked singularity
evolves from regular initial data, however, the proof and even precise
formulation of the conjecture still stands as one of the biggest challenges
in general relativity. Therefore, it seems important to consider possible
astrophysical consequences of the hypothetical existence of naked
singularities. Of particular interest are those effects that could
distinguish a naked singularity from black holes. We shall consider the
influence of a nonzero \cc{} on the character of the simplest, spherically
symmetric naked singularity spacetimes containing a nonzero electric charge
$Q$.

We know a wide class of solutions of the Einstein equations representing
naked singularities. Here, we focus our attention to the simplest class of
solutions of the Einstein\nnd Maxwell equations with $\Lambda \neq 0$ that
describe both naked-singularity and black-hole spacetimes.These solutions
are given by the spherically symmetric \RN\nnd\aodS{} spacetimes.  Using
standard methods, we shall study geodetical motion in these spacetimes and
\ed{} of the central planes of both the ordinary space geometry and optical
reference geometry. The results will be compared with known related
properties of both the \Schw\nnd\aodS{} black-hole spacetimes ($Q=0$) and
\RN{} spacetimes black-hole and naked-singularity spacetimes ($\Lambda=0$).

The paper is organized in the following way. In Section~\ref{class}, the
\RN\nnd\aodS{} spacetimes are separated into the black-hole and
naked-singularity spacetimes in the parameter space. In Section
\ref{motion}, the geodetical motion is discussed for test particles and
photons.  In Section~\ref{escones}, photon escape cones are established
from the properties of the null geodesics. In Section~\ref{ediag}, \ed{} of
the central planes of the ordinary space geometry are constructed and the
limits of embeddability are given. In Section~\ref{opt}, the optical
reference geometry is introduced, and the inertial and gravitational forces
related to the optical geometry are defined and expressed for general
circular motion, i.e., the motion with $r= \text{const}$, $\theta
=\text{const}$.  In Section~\ref{embed}, \ed{} of the central planes of the
optical geometry are constructed. The limits of embeddability are given and
the turning points of the \ed{} are related to the properties of the
centrifugal force. In Section~\ref{asympt}, asymptotic behavior of the
optical geometry near the black-hole horizons and the cosmological horizon
is discussed. In Section~\ref{conrem}, new features of the geodetical
motion and the \ed, caused by the interplay of the electric charge of the
background and a nonzero \cc{} in both the black-hole and naked-singularity
spacetimes, are summarized and briefly discussed.

\section{Classification of the \RN{} spacetimes
         with a nonzero \cc{}}\label{class}
       
In the standard \Schw{} coordinates $(t,r,\theta,\phi)$, and the geometric
units ($c=G=1$), the \RN\nnd\dS{} ($\Lambda > 0$), and \RN\nnd\adS{}
($\Lambda < 0$) spacetimes are given by the line element
\bea
  \diff s^2 =
    &-&\left(
      1-\frac{2M}{r}+\frac{Q^2}{r^2}-\frac{\Lambda}{3}r^2
     \right)\diff t^2
    + \left(
      1-\frac{2M}{r}+\frac{Q^2}{r^2}- \frac{\Lambda}{3}r^2
     \right)^{-1} \diff r^2                                     \nonumber\\
    &+&r^2\left(\diff\theta^2+\sin^2\theta\,\diff\phi^2\right),   \label{e1}
\eea
and the related electromagnetic field is given by the \four potential
\beq                                                              \label{e2}
  A_{\mu} = \frac{Q}{r}\,\delta_{\mu}^t.
\eeq
Here, $M$ denotes mass and $Q$ denotes electric charge of the spacetimes.
However, it is convenient to introduce a dimensionless cosmological
parameter
\beq                                                              \label{e3}
  y \equiv \tfrac{1}{3} \Lambda M^2,
\eeq
a dimensionless charge parameter
\beq                                                              \label{e4}
  e \equiv \frac{Q}{M},
\eeq
and dimensionless coordinates $t \to t/M$, $r \to r/M$.  It is equivalent
to putting $M=1$.

The event horizons of the geometry (\ref{e1}) are determined by the
condition
\beq                                                              \label{e5}
  - g_{tt} \equiv 1- \frac{2}{r}+ \frac{e^2}{r^2} - y{r^2} = 0.
\eeq
The loci of the event horizons can be expressed as solutions of the equation
\beq                                                              \label{e6}
  y= y_{\text{h}}(r;e) \equiv \frac{r^2-2r+ e^2}{r^4}.
\eeq
Inspecting properties of the function $y_{\text{h}} (r;e)$, we determine
distribution of black-hole and naked-singularity spacetimes in the
parameter space ($e^2$-$y$).

In the special case of $e = 0$, the black-hole spacetimes exist for all
$y\leq 0$, and for $0 < y \leq y_{\text{c}} = 1/27$. If $y > 1/27$, a naked
singularity exists, however, there is no static region in these spacetimes
(for details, see Ref.\,\cite{Stu-Hle:1999:PHYSR4:}).

\begin{figure}[b]
\centering
\includegraphics[width=.7\linewidth]{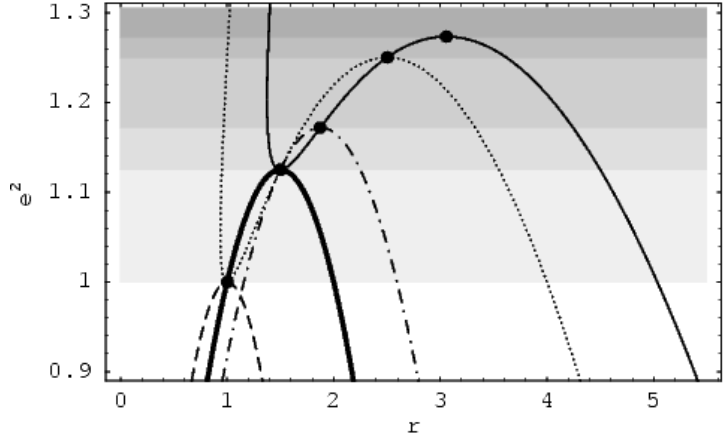}
\caption{The characteristic functions governing the test-particle geodetical
  motion in the \RN\nnd\aodS{} spacetimes. The function
  $e^2_{\text{z(h)}}(r)$ (represented by the dashed curve) governs the
  zero points (event horizons of \RN{} black holes) and the characteristic
  function $e^2_{\text{e(h)}}(r)$ (represented by the bold solid curve)
  governs the local extrema of the function $y_{\text{h}}(r;e)$
  determining the event horizons of the \RN\nnd\aodS{} spacetimes. Their
  respective local maxima are located at the points $(1,1)$ and
  $(3/2,9/8)$.  The characteristic functions $e^2_{\text{z(ms)}\pm}(r)$
  (represented by the dotted curve) govern the zero points of
  $y_{\text{ms}}(r;e)$ determining the marginally stable circular
  geodesics, the characteristic function $e^2_{\text{d(ms)}}(r)$
  (represented by the dashed-dotted curve) governs the divergent points of
  $y_{\text{ms}}(r;e^2)$, and the characteristic functions
  $e^2_{\text{e(ms)}\pm}(r)$ (represented by the thin solid curve) govern
  the local extrema of the function $y_{\text{ms}}(r;e)$.  Their
  respective local maxima are located at the points $(5/2,5/4)$ (on the
  `$-$' branch), $(15/8,75/64)$ and $(55/18,275/216)$ (on the `$-$'
  branch). The local minimum of the function $e^2_{\text{z(ms)}\pm}(r)$
  (the `$-$' branch) coincides with the local maximum of the function
  $e^2_{\text{z(h)}}(r)$ at (1,1), the local minimum of the function
  $e^2_{\text{e(ms)}\pm}(r)$ (the `$-$' branch) coincides with the local
  maximum of the function $e^2_{\text{e(h)}}(r)$ at $(3/2,9/8)$.  The
  extrema of the functions (depicted by bold dots) divide the nonnegative
  region of the parameter $e^2$ into six subintervals emphasized by
  increasing gray levels, each of them implying different behavior of the
  functions $y_{\text{h}}(r;e)$ and/or $y_{\text{ms}}(r;e)$.}
\label{f1}
\end{figure}

\begin{figure}[p]
\centering
\includegraphics[width=.59\linewidth]{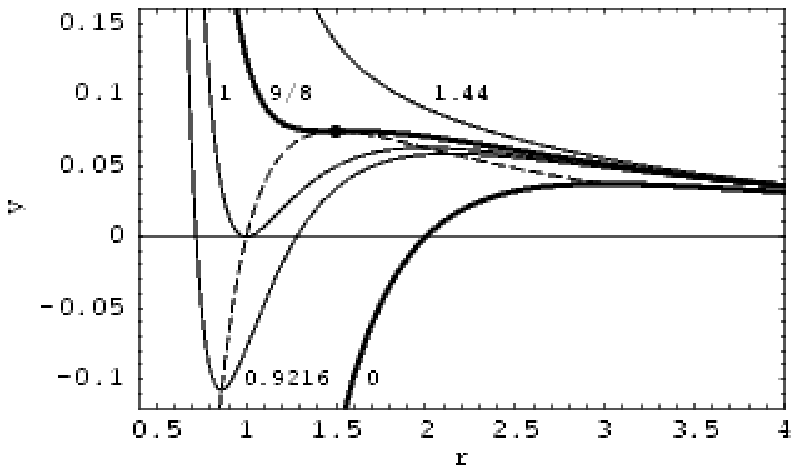}
\caption{Loci of the event horizons in the \RN\nnd\aodS{} spacetimes.
  The functions $y_{\text{h}}(r;e^2)$ depicted in the figure are labeled
  by the corresponding value of the parameter $e^2$. For $e^2=0$, the
  corresponding function (represented by the lower bold curve) diverge to
  $-\infty$ as $r\to 0_+$. Three qualitatively different cases of behavior
  of the function $y_{\text{h}}(r;e^2)$ for $e^2>0$ can be distinguished.
  For $0 < e^2 < 9/8$, one local minimum and one local maximum occur on the
  function $y_{\text{h}}(r;e^2)$, moreover, $y_{\text{h}}(r;e^2) > 0$ if
  $1 < e^2 < 9/8$. For $e^2 = 9/8$, both extrema merge in the inflex point
  (depicted by the bold dot) located at $(3/2,y_{\text{crit}}=2/27)$ on
  the critical curve (represented by the upper bold curve), and for $e^2
  >9/8$ the function $y_{\text{h}}(r;e^2)$ is monotonically decreasing.
  The loci of the local extrema are emphasized by the dashed curve.}
\label{f2}
\end{figure}

\begin{figure}[p]
\centering
\includegraphics[width=.59\linewidth]{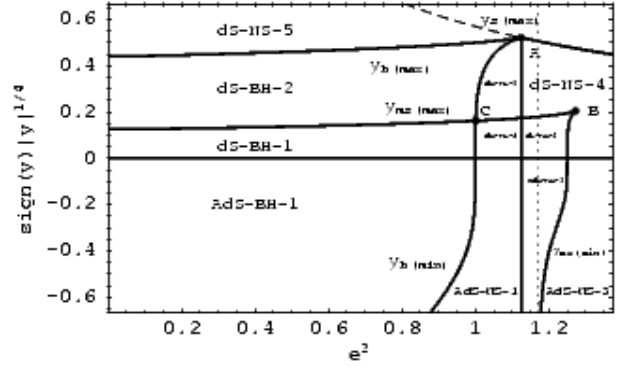}
\caption{The classification of the \RN\nnd\aodS{} spacetimes according
  to properties of the effective potential of the test-particle geodetical
  motion.  The functions $y_{\text{h(min)}}(e^2)$ and
  $y_{\text{h(max)}}(e^2)$ limit the region of black-hole spacetimes
  (shaded) in the parameter space $e^2$-$y$. Outside that region, merely
  naked-singularity spacetimes exist. The function
  $y_{\text{s(max)}}(e^2)$ (represented by the dashed curve in the
  region, where it is irrelevant for the classification) marks the local
  maxima of the function $y_{\text{s}}(r;e^2)$ governing the static
  radii. The functions $y_{\text{ms(min)}}(e)$ and
  $y_{\text{ms(max)}}(e)$ separate the asymptotically \dS{} black-hole
  spacetimes containing a region of stable circular orbits allowing
  accretion processes in the disk regime (dS-BH-1) from those with unstable
  circular orbits only (dS-BH-2), and the naked-singularity spacetimes with
  two regions of stable circular orbits from those with one region of
  stable circular orbits (see the text and Figs~\protect\ref{f5} and \protect\ref{f7}). The
  relevant points have the following coordinates $e^2$ and $y$: $A (9/8,
  2/27)$, $B (275/216,0.001753)$, $C (1.000695,0.0006931)$.  Notice that in
  order to make the distribution of different classes in the parameter
  space more evident, we use the function $\sign(y)|y|^{1/4}$.}
\label{f3}
\end{figure}

For $e^2 > 0$, properties of $y_{\text{h}}(r;e^2)$ can be reviewed in the
following way. The zero points are given by the relation determining
horizons of \RN{} black holes
\beq                                                              \label{e7}
  e^2 = e^2_{\text{z(h)}} \equiv 2r - r^2.
\eeq
There is one divergent point at $r=0$, where $y_{\text{h}}(r\to 0, e^2)
\to +\infty$. The asymptotic behavior is given by $y_{\text{h}}(r \to
\infty, e^2) \to 0$.  The local extrema (given by $\p
y_{\text{h}}(r;e^2)/\p r=0$) are determined by the relation
\beq                                                              \label{e8}
  e^2 = e^2_{\text{e(h)}} \equiv \tfrac{1}{2}(3r-r^2).
\eeq
Both function $e^2_{\text{z(h)}}(r)$, $e^2_{\text{e(h)}}(r)$ are drawn in
Fig.\,\ref{f1}.  At the local extrema, the cosmological parameter is given
by the relations
\bea
  y_{\text{h(max)}}(e) &\equiv&
    \frac{r_{\text{e(h)}+}-e^2}{r^4_{\text{e(h)}+}},           \label{e9}\\
  y_{\text{h(min)}}(e) &\equiv&
    \frac{r_{\text{e(h)}-}-e^2}{r^4_{\text{e(h)}-}},          \label{e10}
\eea
where
\beq                                                            \label {e11}
  r_{\text{e(h)}\pm} = \frac{3}{2}
    \left[1\pm \left(1- \frac{8e^2}{9} \right)^{1/2}\right].
\eeq
Clearly, the local extrema can exist, if $e^2 \leq e^2_{\text{crit}} =
9/8$.  For $e^2 = 9/8$, $y_{\text{h}}(r;e^2)$ has an inflex point at $r=
3/2$, with value of the cosmological parameter
\beq                                                             \label{e12}
  y = y_{\text{crit}} \equiv \tfrac{2}{27}
\eeq
corresponding to the largest value of $y$ that permits the existence of
black-hole solutions. We can distinguish three qualitatively different
cases of behavior of $y_{\text{h}}(r;e^2)$ (see Fig.\,\ref{f2}) according
to the charge parameter $e^2$. Properties of the \RN\nnd\aodS{} spacetimes
relatively to the parameter $y$, $e^2$ can then be classified easily.

If $0<e^2 \leq 1$, there is $y_{\text{h(min)}}<0$; if $1< e^2 \leq 9/8,
y_{\text{h(min)}} >0$; if $e^2 > 9/8$, there are no local extrema.

The black-hole spacetimes (with $e^2 < 9/8$) can exist just if
$y_{\text{h(min)}}(e) < y < y_{\text{h(max)}}(e)$; for $y<
y_{\text{h(min)}}(e)$ or $y> y_{\text{h(max)}}(e)$, the naked-singularity
spacetimes occur.  Clearly, for $e^2 > 9/8$ naked-singularity spacetimes
exist only, while for $1<e^2 < 9/8$ no black holes can exist in
asymptotically \adS{} spacetimes. Both black-hole and naked-singularity
spacetimes in both asymptotically \dS{} and \adS{} universe can exist if
$0< e^2 \leq 1$.  In asymptotically \dS{} spacetimes ($y > 0$), a
black-hole geometry has three horizons $r_{\text{b}-}< r_{\text{b}+}<
r_{\text{c}}$. The geometry is static under the inner black-hole horizon
($0<r<r_{\text{b}-}$), and between the outer black-hole and cosmological
horizons ($r_{\text{b}+} < r <r_{\text{c}}$).  In asymptotically \adS{}
spacetimes ($y<0$), a black hole geometry has two horizons $r_{\text{b}-} <
r_{\text{b}+}$, and the geometry is static under the inner horizon
($0<r<r_{\text{b}-}$) and above the outer horizon ($r>r_{\text{b}+}$).  The
relations (\ref{e9}) and (\ref{e10}) directly yield the distribution of
black-hole and naked-singularity spacetimes in the space of parameters $y$
and $e^2$ (see Fig.\,\ref{f3})

\section{Geodetical motion}\label{motion}

Motion of uncharged test particles and photons is governed by the geodetical
structure of the spacetime. The geodesic equation reads
\beq                                                             \label{e13}
  \frac{\Diff p^{\mu}}{\diff\lambda} = 0,
\eeq
where $p^{\mu}\equiv\diff x^{\mu}/\diff\lambda$ is the \four momentum of
a test particle (photon) and $\lambda$ is the affine parameter related to
the proper time $\tau$ of a test particle by $\tau = \lambda/m$. The
normalization condition reads
\beq                                                             \label{e14}
  p^{\mu}p_{\mu} = - m^2,
\eeq
where $m$ is the rest mass of the particle; $m = 0$ for photons.

It follows from the central symmetry of the geometry (\ref{e1}) that the
geodetical motion is allowed in the central planes only. Due to existence
of the time Killing vector field $\vec{\xi}_{(t)} = \p/\p t$ and the axial
Killing vector field $\vec{\xi}_{(\phi)} = \p/\p \phi$, two constants of
the motion must exist, being the projections of the \four momentum onto the
Killing vectors:
\bea
  p_t      &=& g_{t\mu} p^{\mu}    = - \mathcal{E},           \label{e15}\\
  p_{\phi} &=& g_{\phi\mu} p^{\mu} = \Phi.                      \label{e16}
\eea

In the spacetimes with $\Lambda \neq 0$, the constants of motion
$\mathcal{E}$ and $\Phi$ cannot be interpreted as energy and axial
component of the angular momentum at infinity since the geometry is not
asymptotically flat.  However, it should be interesting to discuss a
possibility to find regions of these spacetimes with character `close' to
the asymptotic regions of the \Schw{} (or \RN) spacetimes.

It is convenient to introduce specific energy $E$, specific axial angular
momentum $L$ and impact parameter $\ell$ by the relations
\beq                                                             \label{e17}
  E=\frac{\mathcal{E}}{m}, \quad
  L= \frac{\Phi}{m}, \quad
  \ell = \frac{\Phi}{\mathcal{E}}.
\eeq
Choosing the plane of the motion to be the equatorial plane ($\theta =
\pi/2$ being constant along the geodesic), we find that the motion of test
particles ($m \neq 0$) can be determined by an `effective potential' of the
radial motion
\beq                                                             \label{e18}
  V^2_{\text{eff}} (r;L,y,e) \equiv
  \left(1- \frac{2}{r}+ \frac{e^2}{r^2} -yr^2 \right)
  \left(1+ \frac{L^2}{r^2} \right).
\eeq
Since
\beq                                                             \label{e19}
  (u^r)^2 = \left(\oder{r}{\tau} \right)^2
          = E^2 - V^2_{\text{eff}} (r;L,y,e),
\eeq
the motion is allowed where
\beq                                                             \label{e20}
  E^2 \geq V^2_{\text{eff}}(r;L,y,e),
\eeq
and the turning points of the radial motion are determined by the condition
$E^2 = V^2_{\text{eff}}(r;L,y,e)$.

The radial motion of photons ($m=0$) is determined by a `generalized
effective potential' $\ell^2_{\text{R}} (r;y,e)$ related to the impact parameter
$\ell$.  The motion is allowed, if
\beq                                                             \label{e21}
  \ell^2 \leq \ell^2_{\text{R}} (r;y,e) \equiv
              \frac{r^4}{r^2-2r+e^2-yr^4},
\eeq
the condition $\ell^2 = \ell^2_{\text{R}} (r,y,e)$ gives the turning points of the
radial motion.

The special case of $e=0$ has been extensively discussed in
Ref.\,\cite{Stu-Hle:1999:PHYSR4:}.  Therefore, we concentrate our
discussion on the case $e^2>0$.  The effective potentials
$V^2_{\text{eff}}(r;L,y,e)$ and $\ell^2_{\text{R}} (r;y,e)$ define turning points of
the radial motion at the static regions of the \RN\nnd\aodS{} spacetimes.
(At the dynamic regions, where the inequalities $V_{\text{eff}}(r;L,y,e)<0$
and $\ell^2_{\text{R}} (r;y,e)<0$ hold, there are no turning points of the radial
motion). $V^2_{\text{eff}}$ is zero at the horizons, while $\ell^2$
diverges there. At $r=0$, $V^2_{\text{eff}} \to + \infty$, while $\ell^2_{\text{R}}
= 0$.  Circular orbits of uncharged test particles correspond to local
extrema of the effective potential ($\p V_{\text{eff}}/\p r = 0$). Maxima
($\p^2 V_{\text{eff}}/\p r^2 < 0$) determine circular orbits unstable with
respect to radial perturbations, minima ($\p^2 V_{\text{eff}}/\p r^2 >0$)
determine stable circular orbits. The specific energy and specific angular
momentum of particles on a circular orbit, at a given $r$, are determined
by the relations
\bea
  E_{\text{c}}(r;y,e)&=&\frac{1-\frac{2}{r}+\frac{e^2}{r^2}-yr^2}%
    {\left(1-\frac{3}{r}+\frac{2e^2}{r^2}\right)^{1/2}},      \label{e22}\\
  L_{\text{c}}(r;y,e) &=& \left(\frac{r-e^2-yr^4}%
    {1-\frac{3}{r}+\frac{2e^2}{r^2}}\right)^{1/2}.              \label{e23}
\eea
(The minus sign for $L_{\text{c}}$ is equivalent to the plus sign in
spherically symmetric spacetimes, therefore, we do not give the minus sign
explicitly here and in the following.)

At $r= r_{\text{ph}+}$, and $r= r_{\text{ph}-}$, where
\beq                                                             \label{e24}
  r_{\text{ph}\pm}(e) = \frac{3}{2}
    \left[1\pm \left(1- \frac{8e^2}{9}\right)^{1/2} \right],
\eeq
both $E_{\text{c}}$ and $L_{\text{c}}$ diverge\Md photon circular orbits
exist at these radii. The photon circular orbits are determined by the
local extrema of the function $\ell^2_{\text{R}}(r;y,e)$, which are located
at $r= r_{\text{ph}\pm}(e)$ independently of the cosmological parameter
$y$. Of course, the impact parameter of the photon circular orbits depends
on $y$; there is
\beq                                                             \label{e25}
  \ell^2_{\text{c}\pm}(y,e) =
    \frac{r^4_{\text{ph}\pm}}%
         {r^2_{\text{ph}\pm}-2r_{\text{ph}\pm}+e^2-yr^4_{\text{ph}\pm}}.
\eeq

The loci of photon circular orbits can be implicitly given by the equation
$e^2 = e^2_{\text{ph}}(r) = e^2_{\text{e(h)}}$.  Because
$r_{\text{ph}\pm}(e) = r_{\text{e(h)}\pm}(e)$, where
$r_{\text{e(h)}\pm}(e)$ determine local extrema of the function
$y_{\text{h}}(r;e)$ governing horizons of the \RN\nnd\aodS{} spacetimes, we
can directly conclude that two photon circular orbits can exist at the
naked-singularity spacetimes with $y< y_{\text{h(min)}}(e)$, while one
photon circular orbit at $r_{\text{ph}+}(e)> r_{\text{b}+}(e)$ exists in
the black-hole spacetimes with $y_{\text{h(min)}}(e)< y<
y_{\text{h(max)}}(e)$. If no local extrema of $y_{\text{h}}(r;e)$ exist,
i.e., for $0<e^2 \leq 9/8$ and $y> y_{\text{h(max)}}(e)$, and for $e^2 >
9/8$ and $y$ arbitrary, no photon circular geodesics are admitted in the
corresponding naked-singularity spacetimes.

The circular geodesics are allowed at regions, where the denominator of
both (\ref{e22}) and (\ref{e23}) is real, i.e., at
\beq                                                             \label{e26}
  r < r_{\text{ph}-} \quad \text{and} \quad  r > r_{\text{ph}+}.
\eeq
However, we have to add the condition given by reality of the numerator in
Eq.\,(\ref{e23}):
\beq                                                             \label{e27}
  r-e^2 - yr^4 \geq 0.
\eeq
The equality at (\ref{e27}) determines so called static radii
$r_{\text{s}}$, where $L_{\text{c}}(r_{\text{s}};y,e) = 0$.

The static radii are given by the condition
\beq                                                             \label{e28}
  y =y_{\text{s}}(r;e) \equiv \frac{r- e^2}{r^4}.
\eeq
The asymptotic behavior of $y_{\text{s}}(r,e)$ is determined by relations
\beq
  y_{\text{s}}(r\to 0, e) \to - \infty, \quad
  y_{\text{s}}(r\to \infty, e) \to 0.
\eeq
The function $y_{\text{s}}(r;e)$ has its zero point at $r=e^2$ and its
local maximum is at $r= 4e^2/3$, where
\beq                                                             \label{e29}
  y_{\text{s(max)}}(e) = \frac{27}{256 e^6}.
\eeq

\begin{figure}[p]
\centering
\includegraphics[width=\linewidth]{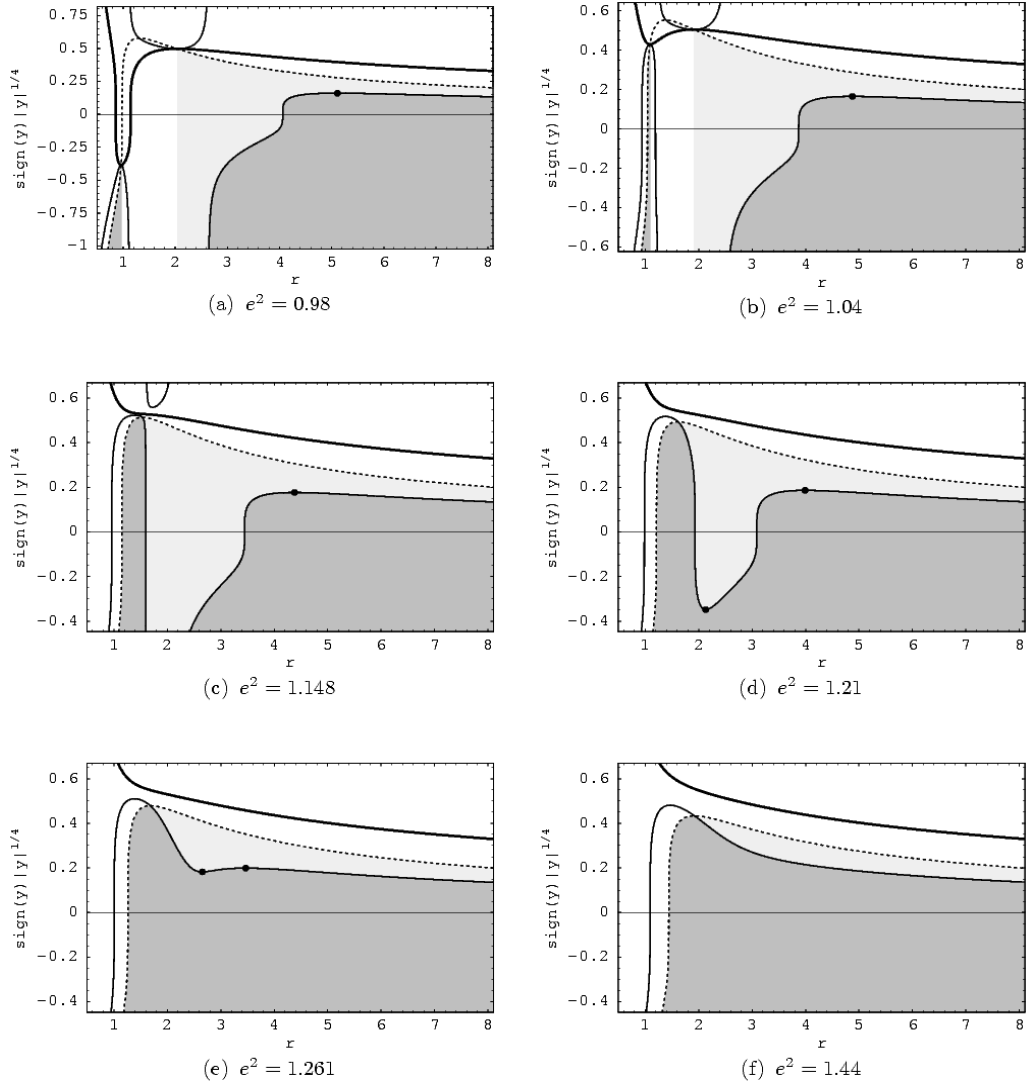}
\caption{Properties of the circular geodesics in the \RN\nnd\aodS{} spacetimes.
  Determined by the functions $y_{\text{h}}(r;e)$ (thick solid lines),
  giving event horizons , $y_{\text{ms}}(r;e)$ (thin solid lines), giving
  marginally stable circular orbits, and $y_{\text{s}}(r;e)$ (dotted
  lines), giving so called static radii, where particles with zero angular
  momentum remain at an equilibrium position. According to the value of the
  charge parameter $e^2$, there are six qualitatively different classes of
  the behavior of the functions $y_{\text{h}}(r;e)$, $y_{\text{ms}}(r;e)$,
  $y_{\text{s}}(r;e)$ (for details see text and Fig.\,\protect\ref{f1}),
  subsequently illustrated in cases (a) through (f). The circular orbits
  are limited by the function $y_{\text{s}}(r;e)$, and by the radii
  $r_{\text{ph}-}$, $r_{\text{ph}+}$, corresponding to photon circular
  orbits existing in the spacetimes with $e^2 < 9/8$ [in cases (a) and
  (b)]. Regions admitting existence of circular geodesics are shaded. If
  the orbits are unstable, a low level of gray is used, while the regions
  of stable circular geodesics are emphasized by a higher level of gray.}
\label{f4}
\end{figure}

The function $y_{\text{s(max)}}(e)$ is illustrated in Fig.\,\ref{f3}.  The
function $y_{\text{s}}(r;e)$ has common points with the function
$y_{\text{h}}(r;e)$ at its extreme points $y_{\text{h(min)}}(e)$, and
$y_{\text{h(max)}}(e)$.  The behavior of $y_{\text{s}}(r)$ is illustrated
in qualitatively different cases in Fig.\,\ref{f4}. The properties of
static radii in dependence on the parameters $e$, $y$ can be summarized in
the following way.

\begin{itemize}
\item[(a)] {\boldmath $0 < e^2 \leq 1$}\enspace\enspace If $y<
  y_{\text{h(min)}}$, there is one static radius at $r_{\text{s}1} <
  r_{\text{ph}-}$.  If $0<y<y_{\text{h(max)}}$, there is one static radius
  at $r_{\text{s}2} > r_{\text{ph}+}$.
\item[(b)] {\boldmath $1 < e^2 \leq 9/8$}\enspace\enspace If $y<0$, there
  is one static radius at $r_{\text{s}1} < r_{\text{ph}-}$. If $0 \leq y <
  y_{\text{h(min)}}$, there are two static radii at $e^2 < r_{\text{s}1} <
  r_{\text{ph}-}$, $r_{\text{s}2} > r_{\text{ph}+}$. If $y_{\text{h(min)}}
  \leq y < y_{\text{h(max)}}$, there is one static radius at
  $r_{\text{s}2} > r_{\text{ph}+}$.
\item[(c)] {\boldmath $e^2 > 9/8$}\enspace\enspace If $y < 0$, there is one
  static radius at $r_{\text{s}1} < 4e^2/3$.  If $0 \leq y <
  y_{\text{s(max)}}$, there are two static radii at $e^2 < r_{\text{s}1} <
  4e^2/3 < r_{\text{s}2}$. If $y=y_{\text{s(max)}}$ there is one static
  radius located at $r_{\text{s}1}=r_{\text{s}2} = 4e^2/3$. If $y >
  y_{\text{s(max)}}$ there are no static radii.
\end{itemize}

If $y > 0$, the gravitational attraction of a black hole acting on a
stationary particle is exactly balanced by the cosmological repulsion at
the static radius; in the case of naked singularities, the balance occurs
at two static radii if $e^2 < 9/8$ and $y < y_{\text{h(min)}}(e)$, or if
$e^2 > 9/8$ and $y < y_{\text{s(max)}}(e)$. On the other hand, static radii
do not exist for naked singularities with $e^2 > 9/8$ and $y >
y_{\text{s(max)}}(e)$. If $y < 0$, there is no static radius in the
black-hole spacetimes, and just one static radius in the naked-singularity
spacetimes. Then the existence of the static radius indicates some
gravitational repulsion connected with the influence of the charge
parameter on the spacetime structure of the \RN\nnd\adS{} naked
singularities. A similar effect occurs in the field of Kerr naked
singularities \cite{deFel:1974:ASTRA:}.

The conditions (\ref{e26}) and (\ref{e27}) limiting radii of circular
geodetical motion have to be considered simultaneously. We arrive at the
conclusion that the geodetical circular orbits are allowed at radii
\beq                                                             \label{e30}
  r_{\text{s}1} < r < r_{\text{ph}-}, \quad
  r_{\text{ph}+} < r <  r_{\text{s}2}.
\eeq

The stable circular geodesics are limited by the relation
\beq                                                             \label{e31}
  4yr^6 - 15yr^5 + 12ye^2r^4 - r^3 + 6r^2 - 9e^2r + 4e^4 \leq 0.
\eeq
Radii of the marginally stable circular geodesics, given by the equality in
(\ref{e31}), can be expressed in the form
\beq                                                             \label{e32}
  y = y_{\text{ms}}(r;e) \equiv
    \frac{r^3 -6r^2 +9e^2r -4e^4}{r^4\left(4r^2 -15r +12e^2 \right)}.
\eeq
The asymptotic behavior of the function $y_{\text{ms}}(r;e)$ is given by
the relations $y_{\text{ms}}(r\to 0, e) \to -\infty$, $y_{\text{ms}}(r \to
\infty, e) \to 0$. The zero points of $y_{\text{ms}}(r;e)$, determining
marginally stable circular geodesics of the \RN{} spacetimes, are given by
the relation
\beq                                                             \label{e33}
  e^2 = e^2_{\text{z(ms)}}(r) \equiv \frac{9r \pm r \sqrt{16r -15}}{8},
\eeq
while its divergent points are located at $r=0$, where $y_{\text{ms}}(r \to
0,e) \to -\infty$, and at radii implicitly determined by the relation
\beq                                                             \label{e34}
  e^2 = e^2_{\text{d(ms)}}(r) \equiv \frac{15r -4r^2}{12}.
\eeq
Both functions $e^2_{\text{z(ms)}}(r)$ and $e^2_{\text{d(ms)}}(r)$ are
illustrated in Fig.\,\ref{f1}. The function $e^2_{\text{z(ms)}}(r)$ has a
local minimum at $r=1$, where $e^2=1$, and a local maximum at $r=5/2$,
where $e^2_{\text{z(ms)(max)}} = 5/4$.  The function
$e^2_{\text{d(ms)}}(r)$ has a local maximum at $r=15/8$, where there is
$e^2_{\text{d(ms)(max)}} = 75/64$.

Local extrema of $y_{\text{ms}}(r;e)$ determine the extremal values
$y_{\text{ms(max)}}(e)$ and $y_{\text{ms(min)}}(e)$ of spacetimes that
admit existence of stable circular geodesics. These local extrema are
determined by the equation
\beq                                                             \label{e35}
  (2e^2 - 3r + r^2)(16e^4 - 28e^2r + 15r^2 - 2r^3) = 0.     
\eeq

In the special case of $e = 0$ , we find $r_{\text{max}} = 15/2$ and
$y_{\text{ms(max)}} = 12/15^4 \approx 0.000237$; further, there is
$r_{\text{min}} = 3$ and $y_{\text{ms(min)}} = 1/27$, which is irrelevant
for timelike geodesics (see Ref.\,\cite{Stu-Hle:1999:PHYSR4:} for details).

In the following we shall restrict our attention to the case of $e^2 > 0$.
The common points of $y_{\text{ms}}(r;e)$ and $y_{\text{h}}(r;e)$ are
located at $r = r_{\text{ph}+}$ and $r = r_{\text{ph}-}$, where both
$y_{\text{ms}}(r;e)$ and $y_{\text{h}}(r;e)$ have local extrema, because of
the first bracket of Eq.\,(\ref{e35}); these are also common points with
the function $y_{\text{s}}(r;e)$. Other local extrema of
$y_{\text{ms}}(r;e)$ are determined by the term in the second bracket in
Eq.\,(\ref{e35}). They can be given by the relation
\beq                                                            \label{e35a}
  e^2 = e^2_{\text{e(ms)}\pm}(r) \equiv \frac{7r - r\sqrt{8r - 11}}{8}.
\eeq

The functions $e^2_{\text{e(ms)}\pm}(r)$ are real at $r \geq 11/8$; at
$r=11/8$ there is $e^2=77/64$. The function $e^2_{\text{e(ms)}-}(r)$ has a
zero point at $r_{\text{z}} = 15/2$ (corresponding to the \Schw\nnd\dS{}
case), a local minimum at $r_{\text{(e}-\text{)min}} = 3/2$, where
$e^2_{\text{e(min)}}= 9/8$, and a local maximum at
$r_{\text{(e}-\text{)max}} = 55/18$, where $e^2_{\text{e(max)}} = 275/216$.
The function $e^2_{\text{e(ms)}\pm}(r)$ is, again, illustrated in
Fig.\,\ref{f1}. Inspecting Fig.\,\ref{f1} we can conclude that the local
extrema of $y_{\text{ms}}(r;e)$, determined by Eq.\,(\ref{e35a}), govern
only one local extreme of $y_{\text{ms}}(r;e)$ in the spacetimes with
black-hole horizons ($e^2 < 9/8$), while they govern three local extrema in
naked-singularity spacetimes with $9/8 < e^2 < 275/216$. Using
Eqs~(\ref{e32}) and (\ref{e35a}), values of the cosmological parameter at
these local extrema, $y_{\text{ms(min)}}(e)$, and $y_{\text{ms(max)}}(e)$
are determined and illustrated in Fig.\,\ref{f3}.

It follows from the discussion presented above that properties of the
circular geodesics in dependence on the charge parameter $e^2$ can be
separated into six qualitatively different cases (see Fig.\,\ref{f1}). They
are illustrated in Fig.\,\ref{f4} giving qualitatively different behavior
of the characteristic functions $y_{\text{h}}(r;e)$, $y_{\text{s}}(r;e)$,
$y_{\text{ms}}(r;e)$; we do not include the special case $e^2 = 0$\Md as it
can be found in Ref.\,\cite{Stu-Hle:1999:PHYSR4:}.  The six qualitatively
different cases of the behavior of the characteristic functions are: (a)~$0
< e^2 \leq 1$, (b)~$1 < e^2 \leq 9/8$ [this interval needs a subdivision
according to the relation of $y_{\text{h(min)}}(e)$, and
$y_{\text{ms(max)}}(e)$; if $e^2 < e^2_{\text{eq}} = 1.000695$, there is
$y_{\text{h(min)}}(e) < y_{\text{ms(max)}}(e) \leq
y_{\text{ms(bh)}}(e^2_{\text{eq}}) = 0.0006931$, while for $e^2 >
e^2_{\text{eq}}$, there is $y_{\text{h(min)}}(e) > y_{\text{ms(max)}}(e)$],
(c)~$9/8 < e^2 < 75/64$, (d)~$75/64 < e^2 < 5/4$, (e)~$5/4 < e^2 <
275/216$, (f)~$e^2 > 275/216$.  The circular geodesics are limited by
relations (\ref{e30}), the lower branch of the circular geodesics
($r_{\text{s}1} < r < r_{\text{ph}-}$) can exist in naked-singularity
spacetimes only. In the black-hole spacetimes no circular photon orbit can
exist under the inner horizon. The stable circular orbits are limited by
$y_{\text{ms}}(r;e)$ from above outside the radii of divergence of this
function. Between the radii of divergence, the stable circular orbits have
to be limited by $y_{\text{ms}}(r;e)$ from below\Md these regions are
always irrelevant for timelike geodesics.

Analysis of the characteristic functions $y_{\text{h}}(r;e)$,
$y_{\text{s}}(r;e)$, $y_{\text{ms}}(r;e)$ shows that there are eleven types
of the \RN\nnd\aodS{} spacetimes with qualitatively different behavior of
the effective potential of the geodetical motion (and the circular orbits).
We shall define the types of the \RN{} spacetimes with a nonzero
cosmological constant according to the properties of the circular
geodesics, as they directly follow from Fig.\,\ref{f4}.

\begin{description}
\item[dS-BH-1] (Type dS-BH-1 means asymptotically \dS{} black-hole
  spacetime of type 1; in the following, the notation has to be read in an
  analogous way.)  One region of circular geodesics at $r > r_{\text{ph}+}$
  with unstable then stable and finally unstable geodesics (for radius
  growing).
\item[dS-BH-2] One region of circular geodesics at $r > r_{\text{ph}+}$
  with unstable geodesics only.
\item[dS-NS-1] Two regions of circular geodesics, the inner region
  consists of stable geodesics only, the outer one contains subsequently
  unstable, then stable and finally unstable circular geodesics.
\item[dS-NS-2] Two regions of circular orbits, the inner one consist of
  stable orbits, the outer one\Md of unstable orbits.
\item[dS-NS-3] One region of circular orbits, subsequently with stable,
  unstable, then stable and finally unstable orbits.
\item[dS-NS-4] One region of circular orbits with stable and then unstable
  orbits.
\item[dS-NS-5] No circular orbits allowed.
\item[AdS-BH-1] One region of circular geodesics at $r > r_{\text{ph}+}$
  with unstable and then stable geo\-des\-ics.
\item[AdS-NS-1] Two regions of circular geodesics, the inner one $(r <
  r_{\text{ph}-})$ consists of stable geodesics only, the outer one $(r >
  r_{\text{ph}+})$ contains both unstable and then stable circular
  geodesics.
\item[AdS-NS-2] One region of circular orbits, subsequently with stable,
  then unstable and finally stable orbits.
\item[AdS-NS-3] One region of circular orbits with stable orbits
  exclusively.
\end{description}

\begin{figure}[p]
\centering
\includegraphics[width=.8\linewidth]{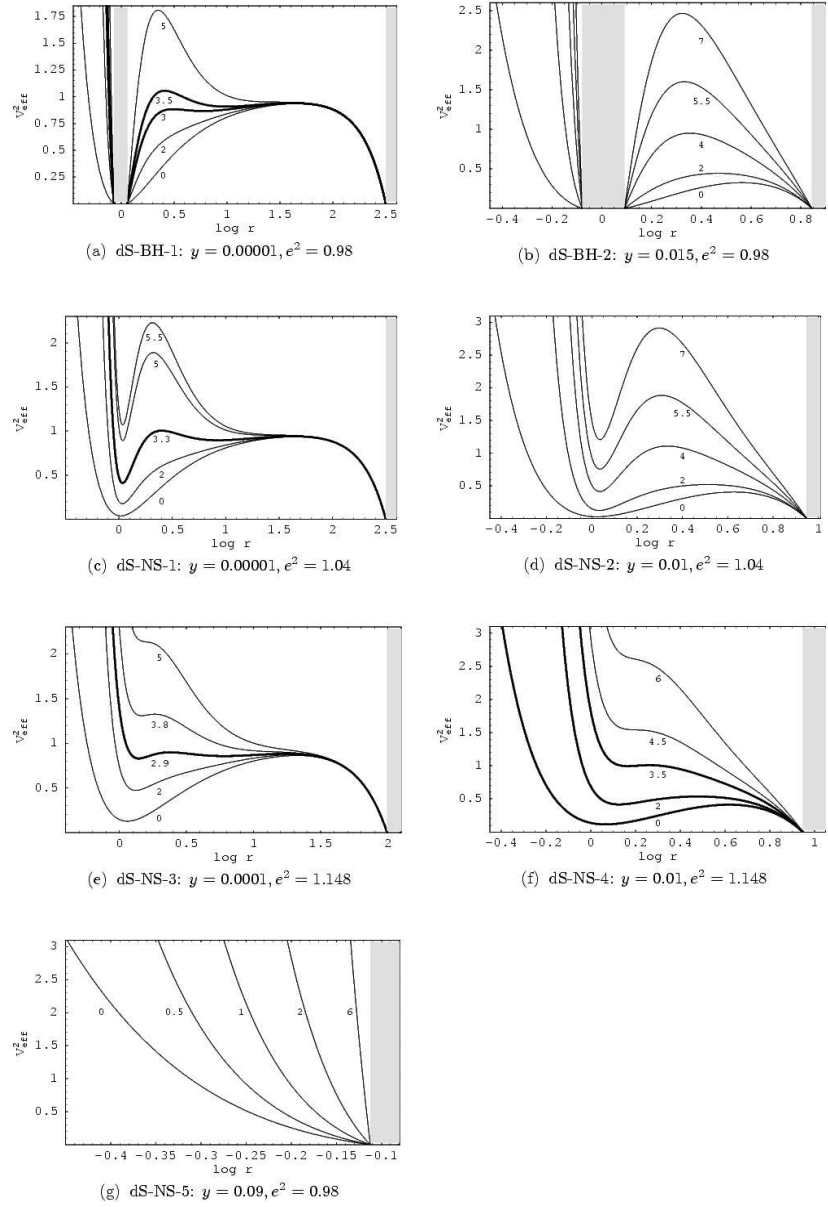}
\caption{The effective potential $V^2_{\text{eff}}(r;e,y,L)$ of the
  test-particle geodetical motion in the asymptotically \dS{} spacetimes.
  There are seven qualitatively different types of the behavior of
  $V^2_{\text{eff}}$ (for details see text and Fig.\,\protect\ref{f4}),
  illustrated in the cases (a) through (g). The dynamic regions of the
  spacetimes, where $V^2_{\text{eff}}$ is not well defined, are shaded. The
  curves representing the effective potential are labeled by values of the
  angular momentum $L$. Note that in the spacetimes dS-NS-1 and 2, the
  character of $V^2_{\text{eff}}$ with the local maxima and minima is
  preserved for $L \to \infty$.}
\label{f5}
\end{figure}

\begin{figure}[p]
\centering
\includegraphics[width=.53\linewidth]{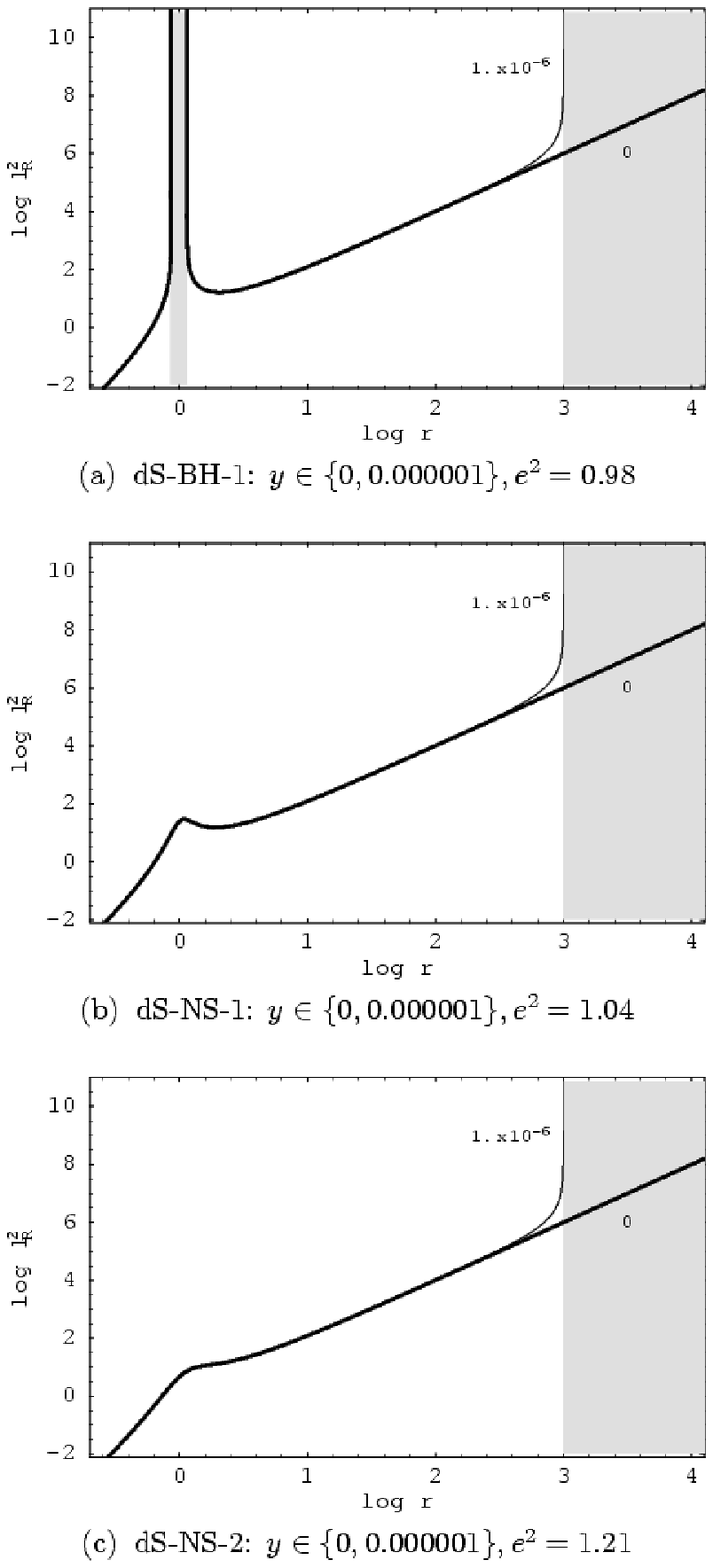}
\caption{The effective potential $\ell^2_{\text{R}}(r;e,y)$ of the photon
  geodetical motion in the asymptotically \dS{} spacetimes. There are three
  qualitatively different types of the behavior of $\ell^2_{\text{R}}$,
  according to the value of the charge parameter $e^2$, as illustrated in
  the cases (a) through (c).  The dynamic regions of the spacetimes, where
  $\ell^2_{\text{R}}$ is not well defined, are shaded. For comparison, the
  effective potential is given for the corresponding \RN{} spacetimes with
  $y=0$ (bold lines). Clearly, the difference occurs at the region near the
  cosmological horizon. Loci of the photon circular orbits (cases (a) and
  (b)) are independent of $y$.}
\label{f6}
\end{figure}

\begin{figure}[t]
\centering
\includegraphics[width=.85\linewidth]{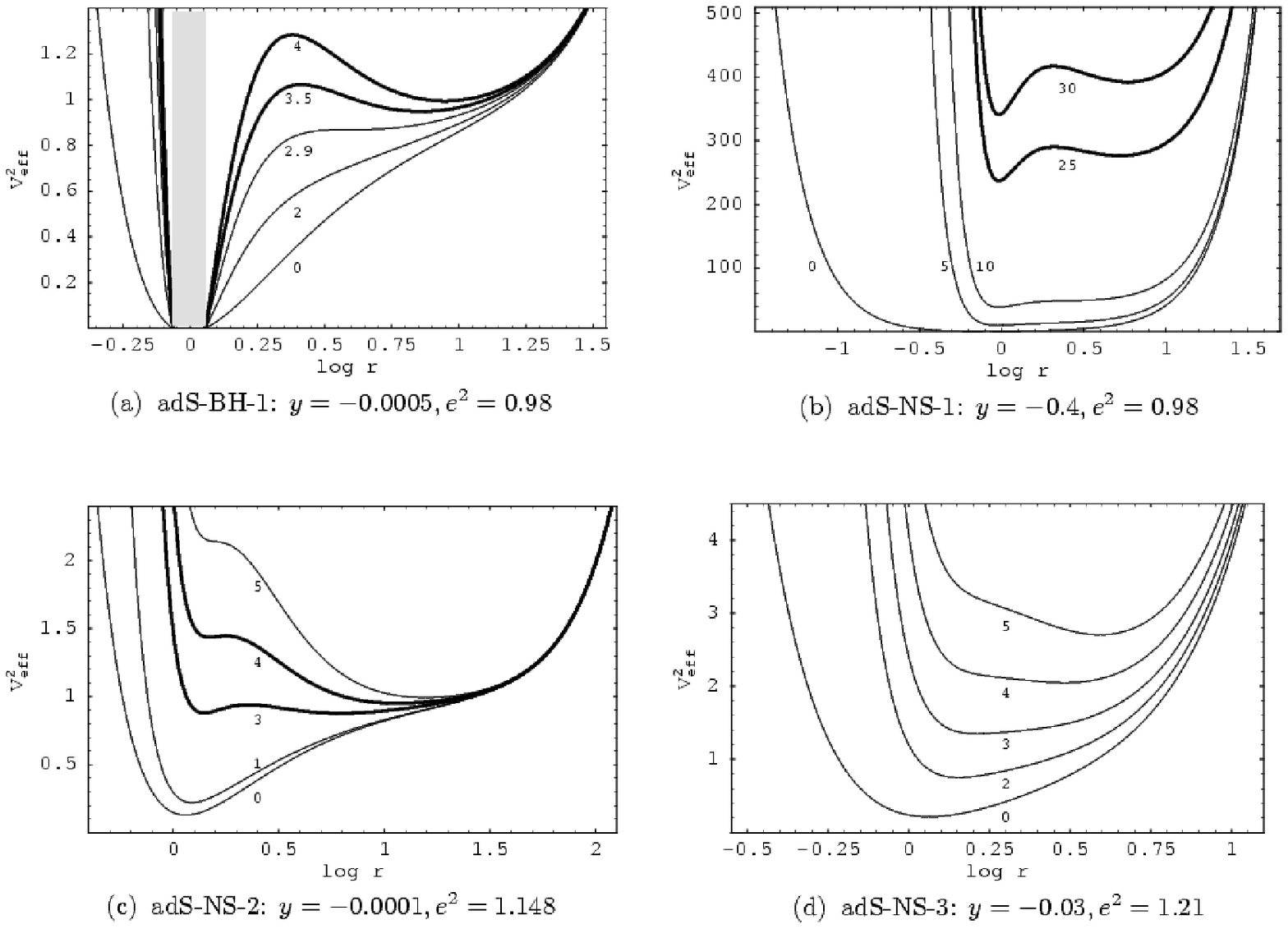}
\caption{The effective potential $V^2_{\text{eff}}(r;e,y,L)$ of the
  test-particle geodetical motion in the asymptotically \adS{} spacetimes.
  There are four qualitatively different types of the behavior of
  $V^2_{\text{eff}}$ subsequently illustrated in the cases (a) through (d).
  The curves giving the effective potential are labeled by the values of
  the particle's angular momentum $L$. In the black-hole spacetimes, only
  one type of the behavior of $V^2_{\text{eff}}$ is possible; it is shown
  in the case a), where the dynamic region of the spacetime is shaded. Note
  that in all of types of the behavior of $V^2_{\text{eff}}$, there is a
  minimum of the effective potential for $L \to \infty$, located at $r \to
  \infty$, with energy of the circular orbit $E_{\text{c}} \to \infty$.
  Therefore, in the limit of ultrarelativistic particles a photon circular
  geodesic appears for $r \to \infty$.}
\label{f7}
\end{figure}

Distribution of these types of the \RN\nnd\aodS{} spacetimes in the
parameter space $e^2$-$y$ is shown in Fig.\,\ref{f3}. According to the
presented classification of the spacetimes, seven (four) types of the
behavior of the effective potential of the geodetical motion in the
asymptotically \dS{} (\adS) spacetimes are shown in Fig.\,\ref{f5}
(Fig.\,\ref{f7}).  Properties of the effective potential can be summarized
in the following way, separating the cases of the asymptotically \dS{} ($y
> 0$), and \adS{} ($y < 0$) spacetimes.

\subsection*{A1\enspace\RN\nnd\dS{} black-hole spacetimes}

There are two types of the black-hole spacetimes (dS-BH-1 and 2) with
different behavior of the effective potential $V^2_{\text{eff}}(r;L,e,y)$
of the test particle motion (see Fig.\,\ref{f5}a,b) In both cases, the
effective potential has no local extrema at $r < r_{\text{b}-}$ for any
value of the specific angular momentum $L$ of the particles. Local minima
of $V^2_{\text{eff}}$, corresponding to stable circular orbits, exist for a
limited range of $L$ in the dS-BH-1 type only. In both dS-BH-1 and 2
spacetimes, there are maxima of $V^2_{\text{eff}}$ at $r > r_{\text{b}+}$.
In dS-BH-1 spacetimes there are even two maxima of $V^2_{\text{eff}}$ in
the range of $L$ allowing the stable circular orbits. One maximum exists
for any $L \geq 0$ in both dS-BH-1 and 2 spacetimes.  In the limit of $L
\to \infty$, its loci $r \to r_{\text{ph}+}$. Therefore, it corresponds to
the photon circular orbit.

The generalized effective potential of the photon motion
$\ell^2_{\text{R}}(r;e,y)$ has the same character for all black-hole
spacetimes (see Fig. 6a). It has always a minimum at $r = r_{\text{ph}} >
r_{\text{b}+}$, corresponding to the photon circular orbit. It diverges at
all the event horizons, and has no local extrema at $r < r_{\text{b}-}$.

We can conclude that both $V^2_{\text{eff}}(r;L,e,y)$ and
$\ell^2_{\text{R}}(r;e,y)$ have the same character as in the case of
\Schw\nnd\dS{} black holes. The differences with respect to the \RN{} black
holes are given by the differences of the asymptotic character of flat and
\dS{} spacetimes.  Especially, there are no \RN{} black holes admitting
unstable circular geodesics only, as happens in the case of dS-BH-2
spacetimes.

\subsection*{A2\enspace\RN\nnd\dS{} naked-singularity spacetimes}

There are five types of the naked-singularity spacetimes (dS-NS-1 through
5), with different behavior of $V^2_{\text{eff}}(r;e,y,L)$ (see
Fig.\,\ref{f5}c--g).  In the spacetimes dS-NS-1 and 2, the effective
potential has a local minimum and a local maximum for all $L \geq 0$. In
the limit of $L \to \infty$, loci of the minimum $r \to r_{\text{ph}-}$,
where $V^2_{\text{eff}}$ diverges, and loci of the maximum $r \to
r_{\text{ph}+}$, where $V^2_{\text{eff}}$ diverges, too. At $r =
r_{\text{ph}-}$ ($r = r_{\text{ph}+}$), the stable (unstable) photon
circular orbit is located. In the dS-NS-1 spacetimes, an additional minimum
and maximum of $V^2_{\text{eff}}$ occurs for a limited range of $L > 0$.
Therefore, two separated regions of both stable and unstable circular
geodesics exist in these spacetimes. In the spacetimes dS-NS-3 and 4, the
local minima and maxima exist for a range of $L$ limited from above ($0
\leq L < L_{\text{u}1}$), and in the dS-NS-3 spacetimes, there is an
additional minimum and maximum for $L$ in the range ($L_{\text{d}} < L <
L_{\text{u}2}$), with $L_{\text{u}2} < L_{\text{u}1}$.  It is important
that in all of the naked\Md singularity spacetimes dS-NS-1 through 4, there
is a stationary particle (a `circular orbit' with $L=0$) at an inner,
stable static radius, and an outer, unstable static radius. An exception is
represented by the spacetimes dS-NS-5 with $V^2_{\text{eff}}$ having no
local extrema.

There are two types of the behavior of the function
$\ell^2_{\text{R}}(r;e,y)$ determining photon motion. For $1 < e^2 <9/8$,
$\ell^2_{\text{R}}$ has an inner local maximum at $r = r_{\text{ph}-}$ (stable
circular photon orbit) and an outer local minimum at $r = r_{\text{ph}+}$
(unstable photon circular orbit)\Md see Fig.\,\ref{f6}b. For $e^2 > 9/8$,
there are no local extrema of $\ell^2_{\text{R}}$ and no photon circular
orbits\Md see Fig.\,\ref{f6}c.

We can conclude that for the photon motion, the new feature of
$\ell^2_{\text{R}}(r;e,y)$ in comparison with the case of \RN{} naked
singularities is given by the asymptotic behavior of $\ell^2_{\text{R}}$
only. On the other hand, for test-particle motion, the differences are
given by the asymptotic behavior in all of the dS-NS-1 through 4
spacetimes, but there are also new features caused by the interplay of the
spacetime parameters $e$ and $y$. In dS-NS-2 through 4 spacetimes, the
inner region of stable orbits is followed by an outer region of unstable
orbits\Md such a behavior is impossible in the field of \RN{} naked
singularities.  Comparison with the \Schw\nnd\dS{} naked-singularity
spacetimes is meaningless as these spacetimes have no static regions.

\subsection*{B1\enspace\RN\nnd\adS{} black-hole spacetimes}

There is only one type of black-hole spacetimes (AdS-BH-1) with respect to
behavior of both $V^2_{\text{eff}}(r;e,y,L)$ and
$\ell^2_{\text{R}}(r;e,y)$.  $V^2_{\text{eff}}$ has a local maximum
(unstable circular orbit) and a local minimum (stable circular orbit) for
the specific angular momentum $L > L_{\text{d}} > 0$.  If $L \to \infty$,
loci of the minimum $r \to r_{\text{ph}+}$ and $V^2_{\text{eff}} \to
\infty$, while loci of the maximum $r \to \infty$ and $V^2_{\text{eff}} \to
\infty$\Md see Fig.\,\ref{f7}a. At $r = r_{\text{ph}+}$, an unstable photon
circular orbit is located; at $r \to\infty$ a stable circular photon orbit
is located, in the sense defined below.

\begin{figure}[p]
\centering
\includegraphics[width=.53\linewidth]{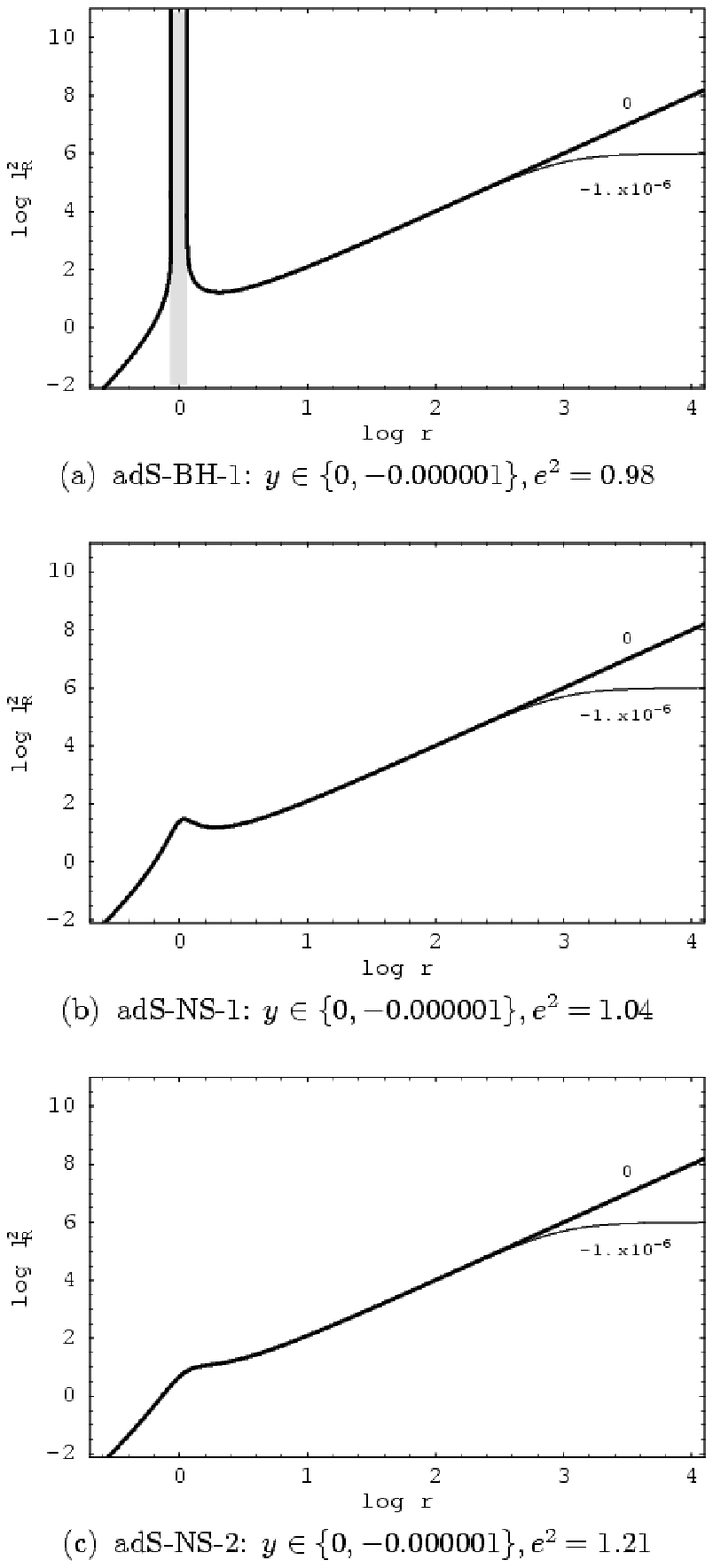}
\caption{The effective potential $\ell^2_{\text{R}}(r;e,y)$ of the photon
  geodetical motion in the asymptotically \adS{} spacetimes. There are
  three qualitatively different types of $\ell^2_{\text{R}}$ according to the
  value of the charge parameter $e^2$, illustrated in the cases (a) through
  (c). The dynamic region of the black-hole spacetime (case (a)) is shaded.
  For comparison, $\ell^2_{\text{R}}(r;e,y=0)$ of the corresponding \RN{}
  spacetimes (drawn by bold solid lines) are included in all of the
  considered cases. Clearly, in all of the cases (a) through (c), the
  qualitative differences are caused by the attractive cosmological
  constant at large radii. Loci of the photon circular orbits, given by
  local extrema of $\ell^2_{\text{R}}$, are independent of $y$.}
\label{f8}
\end{figure}

The generalized effective potential of the photon motion (see
Fig.\,\ref{f8}a) has a local minimum at $r = r_{\text{ph}+}$ for all values
of $e^2 < 1$.  Further, there is
\beq
  \ell^2_{\text{R}}(r\to\infty;e,y) \to (-y)^{-1/2}.           \label{e37b}
\eeq
Therefore, in this sense, a stable circular photon orbit can exist for $r
\to \infty$. It can be considered as a limit of stable circular orbits of
ultrarelativistic particles at radii $r \to \infty$. The function
$\ell^2_{\text{R}}(r;e,y)$ diverges at both event horizons, and has no local
extrema under the inner horizon $r_{\text{b}-}$.

Both $V^2_{\text{eff}}(r;e,y,L)$ and $\ell^2_{\text{R}}(r;e,y)$ have the same
character as in the case of \Schw\nnd\adS{} black holes.  Differences in
comparison with the \RN{} black hole are given by the asymptotic behavior
of $V^2_{\text{eff}}$ and $\ell^2_{\text{R}}$ only.

\subsection*{B2\enspace\RN\nnd\adS{} naked-singularity spacetimes}

There are three types of the naked-singularity spacetimes (AdS-NS-1 through
3) with different behavior of $V^2_{\text{eff}}(r;e,y,L)$\Md see
Fig.\,\ref{f7}b--d. In all of these spacetimes, there is a local minimum of
$V^2_{\text{eff}}$ for $L$ from the interval $0 \leq L < \infty$. If $L=0$,
the minimum corresponds to the stable static radius where a test particle
can be in a stable equilibrium position. If $L > 0$, the minimum
corresponds to a stable circular orbit. For $L \to \infty$, its loci
$r\to\infty$ which corresponds to a stable photon circular orbit in the limit
of stable circular orbits of ultrarelativistic particles, with impact
parameter given by Eq.\,(\ref{e37b}); this is the same situation as in the
case of black-hole spacetimes AdS-BH-1.

In the spacetimes AdS-NS-1, there is additionally a local minimum, and a
local maximum, corresponding to stable and unstable circular orbits, if $0
< L_{\text{d}} < L < \infty$. In the limit of $L \to \infty$ the local
minimum diverges ($V^2_{\text{eff}} \to \infty$) at $r \to r_{\text{ph}-}$,
corresponding to a stable circular photon orbit, while the local maximum
diverges at $r \to r_{\text{ph}+}$, corresponding to an unstable circular
photon orbit. In the spacetimes AdS-NS-2, the additional local minima and
maxima are allowed in a limited range of $L_{\text{d}} < L < L_{\text{u}}$,
while they are not allowed at all in the spacetimes AdS-NS-3.

Similarly to the case of \RN\nnd\dS{} naked singularity spacetimes, there
are two types of the behavior of $\ell^2_{\text{R}}(r;e,y)$ governing
photon motion. For $1 < e^2 < 9/8$ (see Fig.\,\ref{f8}b), there is an inner
local maximum at $r = r_{\text{ph}-}$ (stable circular photon orbit), and
an outer local minimum at $r = r_{\text{ph}+}$ (unstable circular photon
orbit). For $e^2 > 9/8$ (Fig.\,\ref{f8}c), $\ell^2_{\text{R}}(r;e,y)$ has
no local extrema. In both cases, the asymptotic behavior of
$\ell^2_{\text{R}}(r;e,y)$ is given by Eq.\,(\ref{e37b}), as in the
black-hole case.

Comparing behavior of $V^2_{\text{eff}}(r;e,y,L)$ and
$\ell^2_{\text{R}}(r;e,y)$ with the case of \RN{} naked singularities, we
have to emphasize the differences of the behavior in the asymptotic region
of $r \to \infty$.  There is no other difference caused by the interplay of
the charge parameter $e$, and the attractive cosmological parameter $y <
0$.

Distribution of all eleven types of the \RN\nnd\aodS{} spacetimes in the
parameter space $e^2$-$y$ is given in Fig.\,\ref{f3}.

\section{Photon escape cones}\label{escones}

We shall discuss the influence of both the repulsive and attractive \cc{}
on the character of photon escape cones related to the family of static
observers.  Although related to the local observers, the photon escape
cones reflect global properties of the spacetimes under consideration. We
shall consider regions above the outer black-hole horizon and complete
naked-singularity spacetimes.

The orthonormal tetrad of vectors carried by the static observers is
\bea
  \vec{e}_{(t)} &=& \left(1 -\frac{2}{r} +\frac{e^2}{r^2} -yr^2
                \right)^{1/2} \frac{\p}{\p t},                \label{e36}\\
  \vec{e}_{(r)} &=& \left(1 -\frac{2}{r} +\frac{e^2}{r^2} -yr^2
                \right)^{-1/2} \frac{\p}{\p r},               \label{e37}\\
  \vec{e}_{(\theta)} &=& \frac{1}{r} \frac{\p}{\p \theta},    \label{e38}\\
  \vec{e}_{(\phi)} &=& \frac{1}{r \sin \theta}
                                       \frac{\p}{\p \phi}.    \label{e39}
\eea
The components of the \four momentum of a photon as measured by a static
observer are given by
\beq                                                             \label{e40}
  p_{(\alpha)} = p_{\mu} e^{\mu}_{(\alpha)}.
\eeq
Using relations
\beq                                                             \label{e41}
  p^{(t)} = -p_{(t)}, \quad  p^{(\phi)} = p_{(\phi)},
\eeq
we can give the directional angle $\psi$ of the photon, i.e., the angle
measured by the observer relative to its outward radial direction, in the
form
\bea
  \sin\psi &=& \frac{p^{(\phi)}}{p^{(t)}} =
             \left(1- \frac{2}{r} +\frac{e^2}{r^2} -yr^2\right)^{1/2}
             \frac{\ell}{r},                                  \label{e42}\\
  \cos\psi &=& \frac{p^{(r)}}{p^{(t)}} = \pm
                \left[1-\left(1-\frac{2}{r} +\frac{e^2}{r^2} -yr^2
                \right) \frac{\ell^2}{r^2}\right]^{1/2}.         \label{43}
\eea
The photon escape cones can be determined by using the `effective
potential' $\ell^2_{\text{R}} (r;y,e)$ of the photon motion. In establishing the
directional angle $\psi_{\text{c}}$ of a marginally escaping photon, the
impact parameter $\ell_{\text{c}}$ of the unstable circular photon orbit
plays the crucial role for static observers located both under and above
the unstable circular photon orbit at $r=r_{\text{ph}+}(e)$; for simplicity
we consider only positive values of $\ell_{\text{c}}$ because the cone is
symmetric about the radial direction. In appropriately chosen regions of
the naked-singularity spacetimes admitting two circular photon orbits,
there are directional angles corresponding to bound photons.  The bound
photons are concentrated around the stable circular photon orbit at
$r_{\text{ph}-}(e) < r_{\text{ph}+}(e)$.

In the black-hole backgrounds (above their outer horizon), the photon
directional angles can be separated into the escape cones, and the
complementary captured cones. In the naked-singularity backgrounds
admitting two circular photon orbits, there is a region extending under
the radius of the unstable photon circular orbit and crossing the radius of
the stable photon circular orbit, where the escape cone is complemented by
the cone corresponding to directions of bound photons. This bound cone is
concentrated about the direction corresponding to the stable photon
circular orbit.

At $r= r_{\text{ph}+}(e)$, the escape angle is determined by
\beq                                                             \label{e44}
  \psi_{\text{c}}(r=r_{\text{ph}+},y,e) = \frac{\pi}{2},
\eeq
independently of $y$. At $r \neq r_{\text{ph}+}(e)$, escaping directional
angles are determined by the formulae
\bea
  \sin\psi_{\text{c}}(r;y,e) &=& \left(1-\frac{2}{r} +\frac{e^2}{r^2}
    - yr^2 \right)^{1/2} \frac{\ell_{\text{c}+}(y,e)}{r},     \label{e45}\\
  \cos\psi_{\text{c}}(r;y,e) &=& \pm \left[1- \left(1- \frac{2}{r} +
    \frac{e^2}{r^2} -yr^2 \right)
    \frac{\ell^2_{\text{c}+}(y,e)}{r^2} \right]^{1/2},        \label{e46}
\eea
where $\ell_{\text{c}+}(y,e)$ is given by Eq.\,(\ref{e25}). For
$r<r_{\text{ph}+}$, the marginally escaping photon is radially outwards
directed ($p^{(r)} >0$ and $\cos\psi_{\text{c}}$ is taken with `$+$' sign),
for $r> r_{\text{ph}+}$, it is inwards directed ($p^{(r)} <0$ and $\cos
\psi_{\text{c}}$ is taken with `$-$' sign). If $e^2 =0$, and $y \neq 0$, we
arrive at the relations for the escape directional angles at the
\Schw\nnd\aodS{} spacetimes \cite{Stu-Hle:1999:PHYSR4:}
\bea
  \sin\psi_{\text{c}}(r;y) &=& \left[\frac{27(r -2 -yr^3)}
                {r^3(1 -27y)} \right]^{1/2},                  \label{e47}\\
  \cos\psi_{\text{c}}(r;y) &=& \pm \left[\frac{r^3 -27r +54}
                {r^3(1 -27y)} \right]^{1/2}.                    \label{e48}
\eea

\begin{figure}[p]
\centering
\includegraphics[width=.97\linewidth]{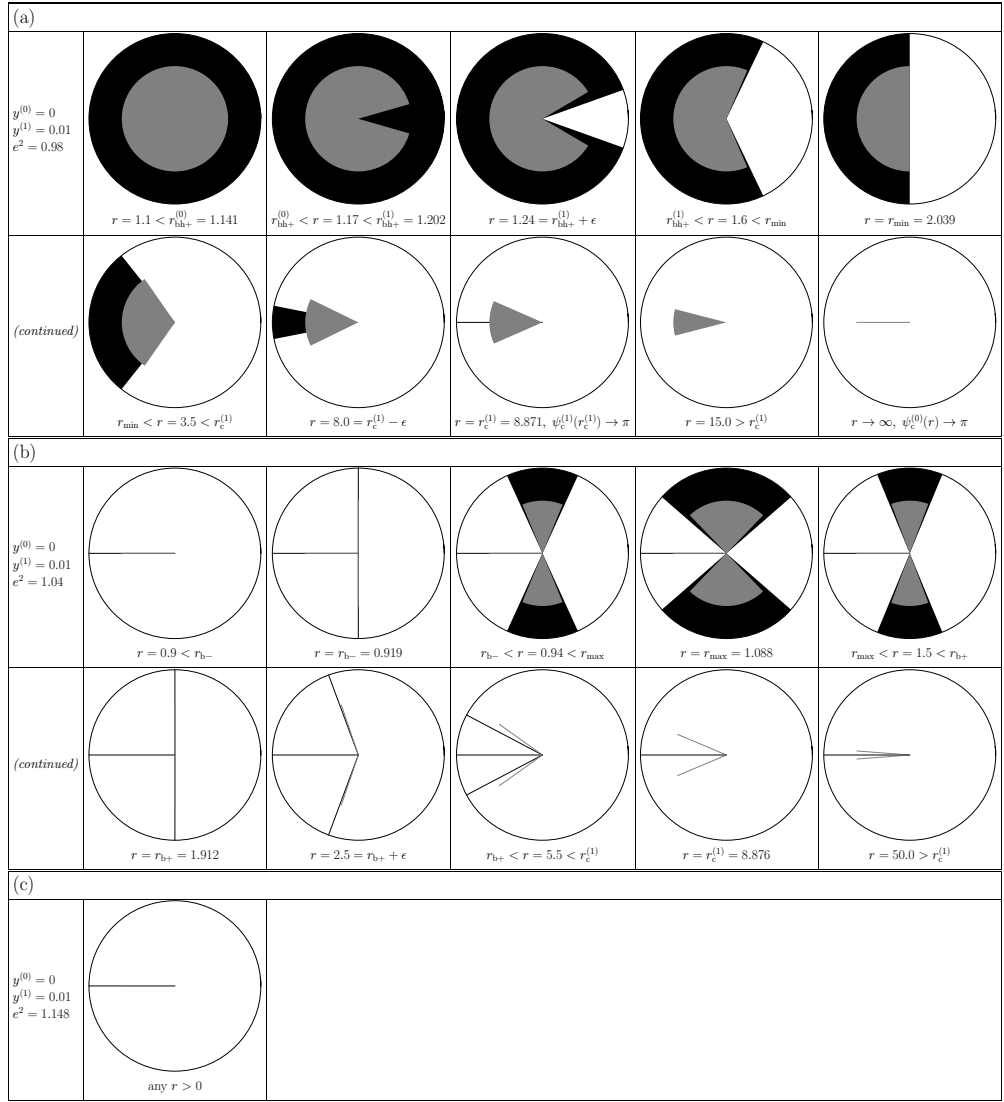}
\caption{The photon
  escape cones in the asymptotically \dS{} spacetimes. They are constructed
  at a representative sequence of radii. There are three qualitatively
  different types of the character of the escape cones, in accord with
  behaviour of $\ell^2_{\text{R}}(r;e,y)$ (cf.\, Fig.\,\protect\ref{f6}).
  Complementary to the escape cones are the capture cones in the black-hole
  spacetimes (case (a)), and the bound cones in the naked-singularity
  spacetimes (case (b)). Because of the central symmetry, we show central
  sections of the cones. The inward-directed radial direction is
  represented by the horizontal, left-oriented line from the centre. The
  captured cones (case (a)) and bound cones (case (b)) are dark shaded. For
  comparison, we consider the escape and the complementary cones in the
  corresponding \RN{} spacetimes; in this case the captured or bound cones
  are gray shaded. Notice that at all radii in the naked-singularity
  spacetimes, the captured photons correspond to a singular case of inward
  radial direction.}
\label{f9}
\end{figure}

\begin{figure}[p]
\centering
\includegraphics[width=.9\linewidth]{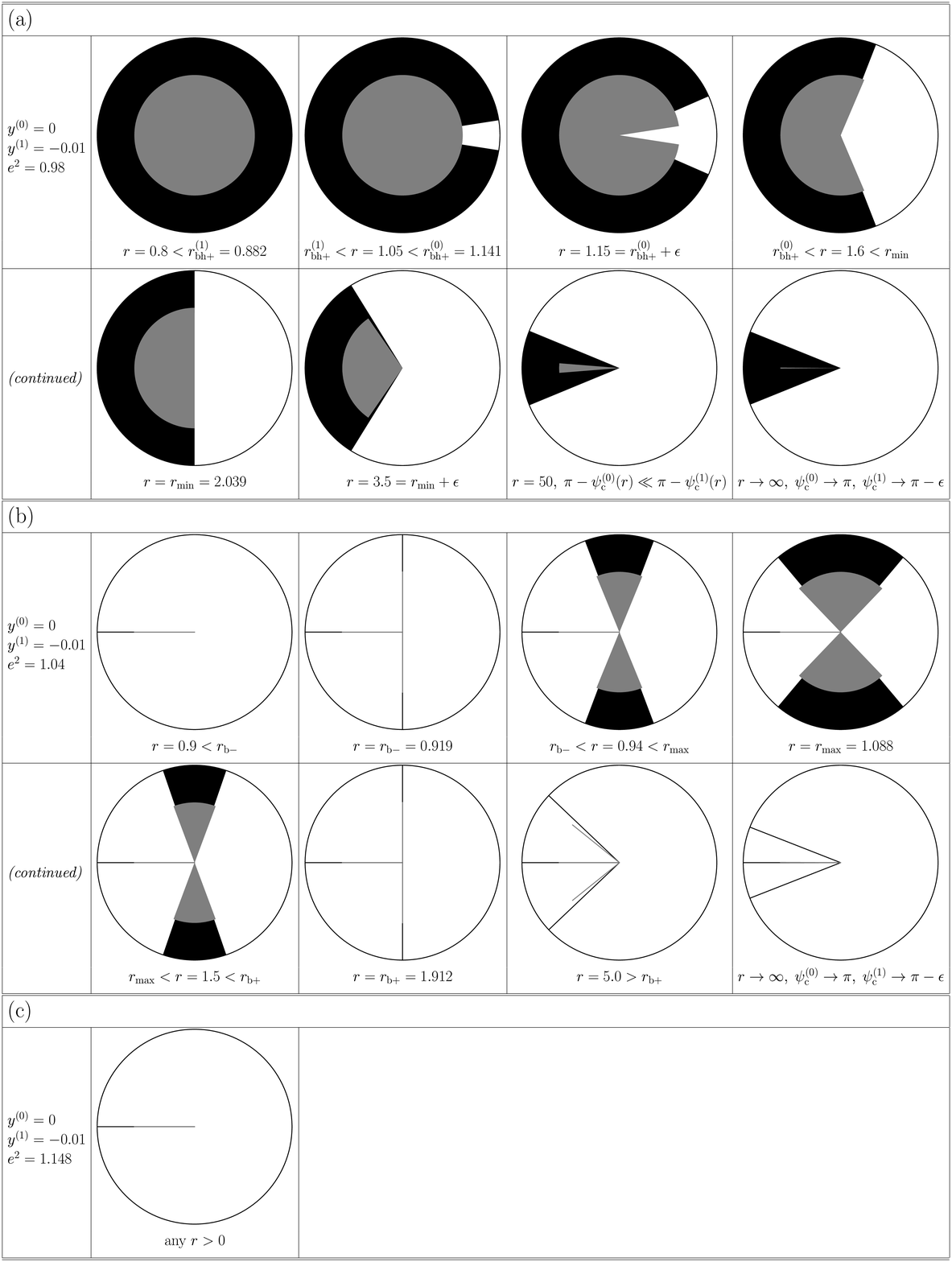}
\caption{The photon escape
  cones in the asymptotically \adS{} spacetimes. Constructed at
  representatively selected radii. There are three qualitatively different
  types of the behaviour of the escape cones, in accord with behaviour of
  $\ell^2_{\text{R}}(r;e,y)$ (cf.\,Fig.\,\protect\ref{f8}). The cones are constructed
  in the same way as for the asymptotically \dS{} spacetimes (see the
  caption of Fig.\,\protect\ref{f9}). Notice that due to the influence of the
  attractive cosmological constant, the capture cones at the black-hole
  spacetimes remain nonzero in the limit of $r \to \infty$, and their
  extension is determined by the cosmological parameter $y$.}
\label{f10}
\end{figure}

Using the relations for the escaping directional angles, the behavior of
the escape cones is established and illustrated in Fig.\,\ref{f9}
(Fig.\,\ref{f10}) for asymptotically \dS{} (\adS{}) spacetimes. For
comparison, the escape cones corresponding to the \RN{} spacetimes with the
same parameter $e$ and $y=0$ are included. The results are summarized in
the following way.

\subsection*{A1\enspace The \RN\nnd\dS{} black-hole spacetimes}



The behavior of the escape cones is presented in Fig.\,\ref{f9}a. At a
fixed (and allowed) $r< r_{\text{ph}+}$, the escape cone of the \RN{} black
hole is the widest one, and it gets smaller with $y$ growing. On the other
hand, at a fixed (and allowed) $r> r_{\text{ph}+}$, the \RN{} escape cone
is smallest one. Of course, the complementary \RN{} photon capture cone is
the widest one at $r> r_{\text{ph}+}$. Close to the cosmological horizon,
the \RN\nnd\dS{} capture cone gets to be strongly narrower than the \RN{}
cone.

\subsection*{A2\enspace The \RN\nnd\dS{} naked-singularity spacetimes}

The behavior of the escape and bound cones is presented in
Fig.\,\ref{f9}b. In the spacetimes with $e^2 < 9/8$, the bound cones are
restricted to some region under the outer, unstable circular photon orbit
at $r_{\text{ph}+}$.  They are concentrated around the inner, stable
circular photon orbit at $r_{\text{ph}-}$, where the bound cones are most
extended. At $r > r_{\text{ph}+}$, there are singular directions
corresponding to bound photons, radiated at the $\psi_{\text{c}}$
directional angle in the inward direction, which wind up at the unstable
photon circular orbit.  For $r \to r_{\text{c}}$, there is $\psi_{\text{c}}
\to \pi$.  For the spacetimes with $e^2 > 9/8$, there are only escaping
photons at all radii up to the cosmological horizon (Fig.\,\ref{f9}c). In
both types of the naked-singularity spacetimes, at all radii $r <
r_{\text{c}}$ the captured cones degenerate to the singular case of photons
with $l = 0$ and $\psi = \pi$, terminated at the physical singularity
($r=0$).

\subsection*{B1\enspace The \RN\nnd\adS{} black-hole spacetimes}


The behavior of the escape (capture) cones is presented in
Fig.\,\ref{f10}a.  At a fixed (and allowed) $r<r_{\text{ph}+}$, the escape
cone of the \RN{} spacetime is the narrowest one; if gets wider with $y$
descending. At a fixed $r>r_{\text{ph}+}$, the \RN{} escape cone becomes
wider than the cones with $y<0$.  The complementary photon capture cone of
the \RN{} spacetime lies inside the capture cones of the spacetimes with
$y<0$.  Asymptotically ($r \to \infty$), the \RN{} capture cone degenerates
into the inward radial direction, while the \RN\nnd\adS{} cone converges to
a cone with a nonzero opening angle equal to
\beq                                                             \label{e49}
  \psi_{\text{c}}(r\to \infty, y<0, e) \sim [\sin^{-1}
             \left(\ell_{\text{c}+}(y,e)\right)(-y)^{1/2}].  
\eeq

Now, the asymptotic behavior of the capture cones is strongly influenced
by the asymptotic character of these spacetimes.

\subsection*{B2\enspace The \RN\nnd\adS{} naked-singularity spacetimes}

For the spacetimes with $e^2 < 9/8$, the behavior of the escape and bound
cones is presented in Fig.\,\ref{f10}b.  The bound cones are widest for the
\RN{} naked singularities. At $r > r_{\text{ph}+}$, there is again the
singular direction corresponding to bound photons which wind up on the
unstable photon orbit.  However, for $r \to \infty$, there is $\pi -
\psi_{\text{c}} \neq 0$. The angle $\psi_{\text{c}}$ is given by
Eq.~(\ref{e49}). If $e^2 > 9/8$, all photons escape from any $r > 0$. In
both types of the naked-singularity spacetimes the capture cones shrink at
all radii $r > 0$ to the singular case of photons radiated radially inwards
with $l=0$ and $\psi = \pi$ that terminate at the singularity at $r=0$.

\section{Embedding of the ordinary geometry}\label{ediag}

Curvature of the static parts of the black-hole and naked-singularity \RN{}
spacetimes with a nonzero \cc{} can be represented in highly illustrative
way by embedding diagrams. Comparison of these embedding diagrams with
those constructed for the pure \RN{} or \Schw{} spacetimes
\cite{Mis-Tho-Whe:1973:Gra:,Kri-Son-Abr:1998:GENRG2:} gives an intuitive
insight into the change of the character of the spacetime caused by the
\cc{} and its interplay with the charge parameter of the spacetime.

The existence of the field of the time Killing vector $\p/\p t$ enables us
to define a privileged notion of space using the hypersurfaces of $t =
\text{const}$. We shall use the induced metric on these hypersurfaces
(i.e., the space components of the metric tensor $g_{ik}$)\Md we call it
ordinary space. Geometry of all the central planes of the ordinary geometry
is the same as those of the equatorial plane ($\theta=\pi/2 =
\text{const}$) as a consequence of the central symmetry of the spherically
symmetric spacetimes.  Therefore, the embedding diagrams will be
constructed for the simplest case of the equatorial plane.

We shall embed the surface described by the line element
\beq                                                             \label{e50}
  \diff\ell^2_{\text{(RNdS)}} =
                \left(1- \frac{2}{r} +\frac{e^2}{r^2} -yr^2 \right)^{-1}
                \diff r^2 +r^2\,\diff\phi^2
\eeq
into the flat Euclidean three-dimensional space with line element expressed
in the standard cylindrical coordinates ($\rho,z,\phi$) in the form
\beq                                                             \label{e51}
  \diff \sigma^2 = \diff \rho^2 +\rho^2\,\diff \phi^2 +\diff z^2.
\eeq
The embedding is represented by the rotationally symmetric surface
$z=z(\rho)$ in the Euclidean space. Its geometry has to be isometric with
the geometry of the equatorial plane described by (\ref{e50}). Thus we have
to identify the line element
\beq                                                             \label{e52}
  \diff \ell^2_{\text{(E)}} =
    \left[1+ \left(\frac{\diff z}{\diff \rho}\right)^2
                \right] \diff\rho^2 +\rho^2\,\diff\phi^2,
\eeq
with the line element (\ref{e50}).  We can identify both the azimuthal and
radial coordinates ($r \equiv \rho$). The embedding diagram can then be
given by the embedding formula $z = z(r)$, which can be obtained by
integrating the relation
\beq                                                             \label{e53}
  \frac{\diff z}{\diff r} = \pm
                \left(\frac{yr^4 +2r -e^2}
                {-yr^4 +r^2 -2r +e^2} \right)^{1/2}.
\eeq
The sign in this formula is irrelevant, leading to isometric surfaces.

\begin{figure}[t]
\centering
\includegraphics[width=.7\linewidth]{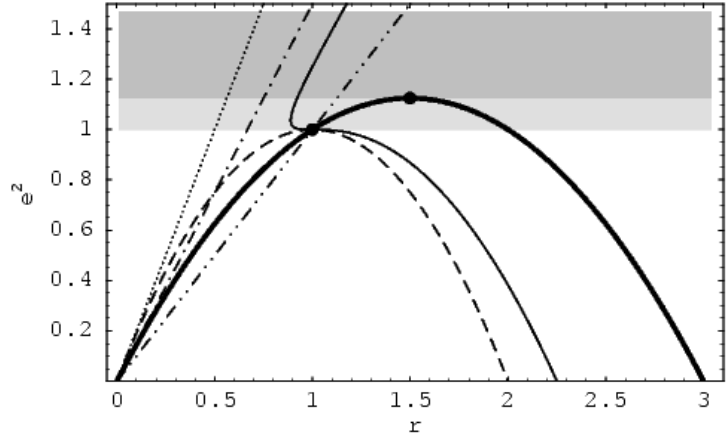}
\caption{The characteristic
  functions governing properties of the embedding diagrams of both the
  ordinary space and the optical reference geometry associated with the
  \RN\nnd\aodS{} spacetimes. The characteristic function
  $e^2_{\text{e(h)}}(r)$ (represented by the bold solid line) governs the
  local extrema of the function $y_{\text{h}}(r;e)$ determining the event
  horizons of the spacetimes. Its local maximum is located at the point
  $(3/2, 9/8)$. The function $e^2_{\text{z(h)}}(r)$ (represented by the
  dashed line) governs the zero points of $y_{\text{h}}(r;e)$, i.e., the
  event horizons of the \RN{} black holes. The function
  $e^2_{\text{z(e(ord))}}(r)$ (dotted line) represents zero points and the
  function $e^2_{\text{min(e(ord))}}(r)$ (dash-dotted line) represents the
  local minimum of the function $y_{\text{e(ord)}}(r;e)$ governing the
  limits of embeddability of the ordinary space. The function
  $e^2_{\text{z(e(opt))}}(r)$ (light solid line) determines the zero point
  of the function $y_{\text{e(opt)}}(r;e)$ governing the limits of
  embeddability of the optical space. The local extrema of
  $y_{\text{e(opt)}}(r;e)$ are given by the function $e^2=r$
  (dash-two-dotted line), and by the function $e^2_{\text{e(h)}}(r)$ (bold
  solid line) governing also the radii of photon circular geodesics that
  are independent of the cosmological parameter $y$, coinciding with the
  extreme points of $y_{\text{h}}(r;e)$. The extrema of the functions
  $e^2_{\text{e(h)}}(r)$ and $e^2_{\text{z(h)}}(r)$ (depicted by the bold
  dots) divide the nonnegative region of the charge parameter $e^2$ into
  three subintervals emphasized by increasing gray level, each of them
  implying different behavior of the functions $y_{\text{h}}(r;e)$,
  $y_{\text{e(ord)}}(r;e)$, $y_{\text{e(opt)}}(r;e)$.}
\label{f11}
\end{figure}

\begin{figure}[t]
\centering
\includegraphics[width=.7\linewidth]{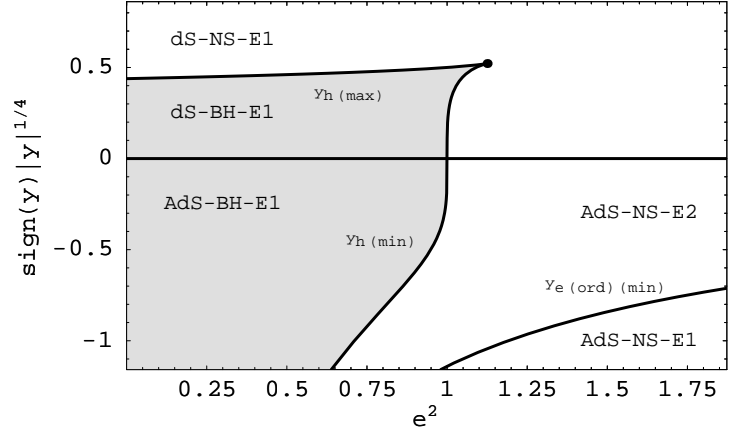}
\caption{The
  classification of the \RN\nnd\aodS{} spacetimes according to the
  properties of the embedding diagrams of their ordinary space geometry.
  The functions $y_{\text{h(min)}}(e)$ and $y_{\text{h(max)}}(e)$ limit the
  black-hole spacetimes (shaded region) in the parameter space $e^2$-$y$.
  Outside that region, merely naked-singularity spacetimes exist. The
  function $y_{\text{e(ord)(min)}}(e)$ separates the spacetimes AdS-NS-E1
  having no space region embeddable into the Euclidean space.}
\label{f12}
\end{figure}

Since the embedding diagrams have to be constructed at the static regions,
where denominator at the right hand side of (\ref{e53}) is nonzero, the
limits of embeddability are given by the condition
\beq                                                             \label{e54}
  yr^4 +2r -e^2 \geq 0,
\eeq
which can be rewritten in the form
\beq                                                             \label{e55}
  y \geq y_{\text{e(ord)}}(r;e) \equiv \frac{-2r + e^2}{r^4}.
\eeq
The asymptotic behavior of the function $y_{\text{e(ord)}}(r;e)$
determining limit of embeddability of the ordinary space is given by
relations
\beq
  y_{\text{e(ord)}}(r \to 0, e) \to + \infty, \quad
  y_{\text{e(ord)}}(r \to \infty, e) \to 0.
\eeq
The zero point of the function $y_{\text{e(ord)}}(r;e)$ is given by
\beq                                                             \label{e56}
  e^2 = e^2_{\text{z(e(ord))}} \equiv 2r,
\eeq
its local minimum is located at
\beq                                                             \label{e57}
  e^2 = e^2_{\text{min(e(ord))}} = \tfrac{3}{2}r,
\eeq
where 
\beq                                                             \label{e58}
  y_{\text{e(ord)(min)}}(e) = - \frac{27}{16e^6}.
\eeq
The functions (\ref{e56}) and (\ref{e57}) are illustrated in
Fig.\,\ref{f11}. The function (\ref{e58}) is illustrated in
Fig.\,\ref{f12}.


\begin{figure}[p]
\centering
\includegraphics[width=.59\linewidth]{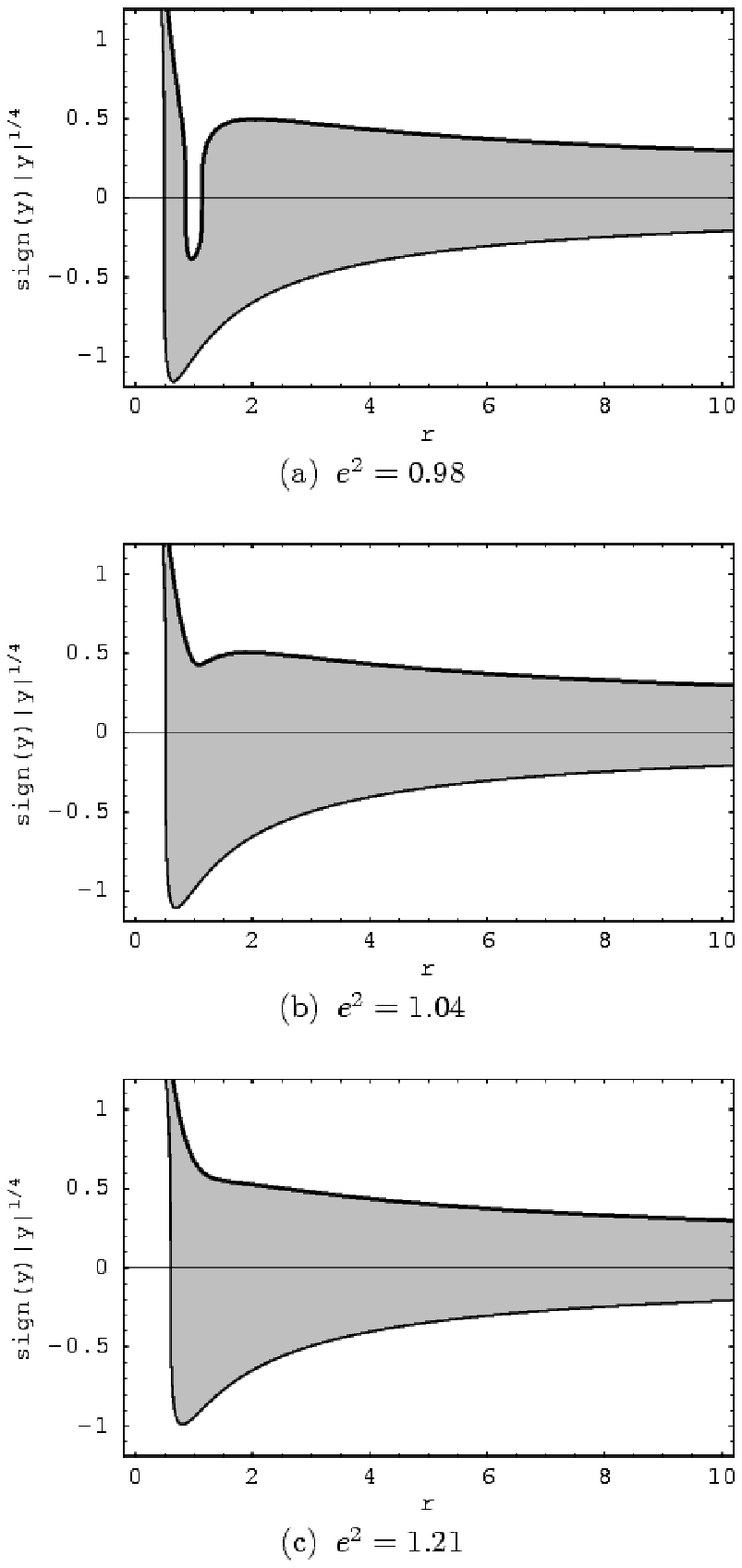}
\caption{Embeddability of
  the ordinary space geometry of the \RN\nnd\aodS{} spacetimes. According
  to values of the charge parameter $e^2$, there are three qualitatively
  different types (cases (a) through (c)) of the behaviour of the functions
  $y_{\text{h}}(r;e)$ (bold solid lines), and $y_{\text{e(ord)}}(r;e)$
  (thin solid lines) determining regions of embeddability. The regions of
  embeddability are shaded.}
\label{f13}
\end{figure}

The regions of embeddability are given by the condition
\beq                                                             \label{e59}
  y_{\text{e(ord)}}(r;e) < y < y_{\text{h}}(r;e).
\eeq                                            
The embeddings of the \Schw\nnd\aodS{} spacetimes are discussed in
Ref.\,\cite{Stu-Hle:1999:PHYSR4:} and will not be repeated here.  There are
three qualitatively different types of the behavior of the functions
$y_{\text{e(ord)}}(r;e)$ and $y_{\text{h}}(r;e)$ in the \RN{} spacetimes
with $\Lambda \neq 0$: (a)~$0 < e^2 \leq 1$, (b)~$1 < e^2 \leq 9/8$,
(c)~$e^2 > 9/8$.  They are illustrated in Fig.\,\ref{f13}.

Analysis of these functions enables an appropriate classification of the
spacetimes according to the properties of the embedding diagrams of the
ordinary space. There are five types of the \RN\nnd\aodS{} spacetimes with
qualitatively different behavior of the embedding diagrams as directly
follows from Fig.\,\ref{f13}.

{\sloppy
In the parameter space $e^2$-$y$, distribution of the five different types
of the \RN\nnd\aodS{} spacetimes according to properties of the embedding
diagrams of their ordinary (space) geometry is quite simple, as it follows
distribution of black holes and naked singularities with $y>0$ and $y<0$ .
There is the only exception represented by asymptotically \adS{} naked
singularities which are separated into two parts (AdS-NS-1 and 2), as shown
in Fig.\,\ref{f12}.
\par}

We briefly summarize the properties of the embedding diagrams.

\begin{figure}[p]
\centering
\includegraphics[width=\linewidth]{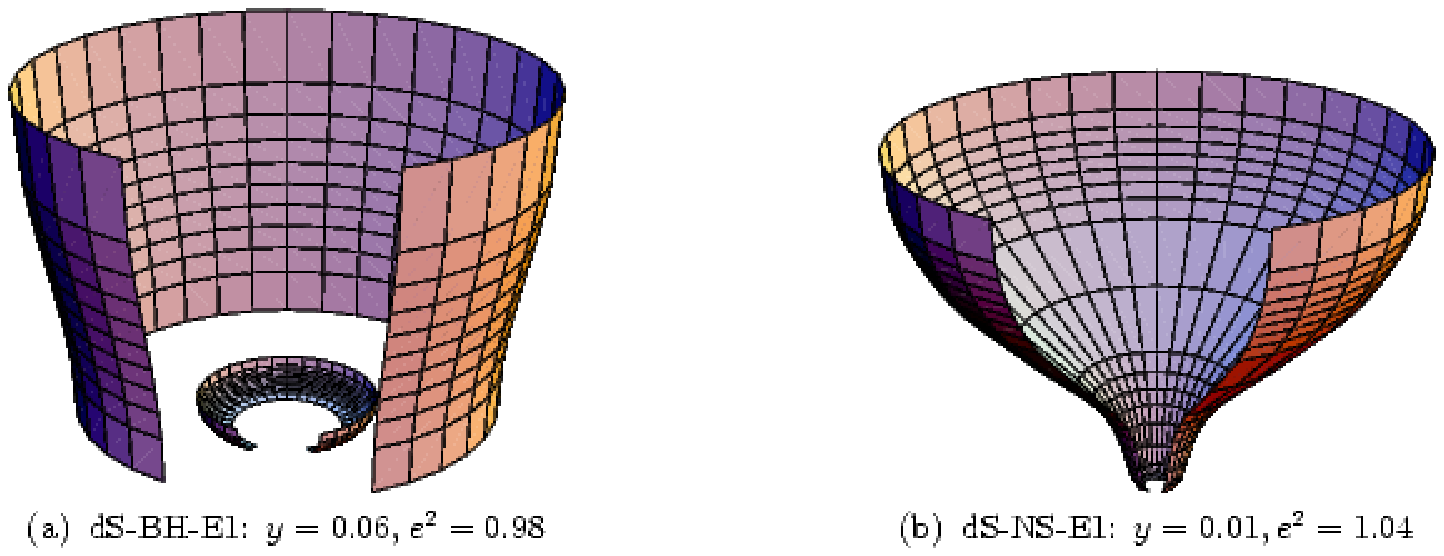}
\caption{Typical embedding
  diagrams of the ordinary space geometry of the asymptotically \dS{}
  spacetimes. There is one type of the embedding diagrams for black holes
  (case (a)), and the other one for naked singularities (case b). In both
  cases there is a funnel near the cosmological horizon. For black holes,
  the embedding can be extended down to the outer horizon and up to the
  inner horizon. For both black holes and naked singularities, the
  embedding cannot be extended down to the singularity at $r=0$.}
\label{f14}
\end{figure} 
\begin{figure}[p]
\centering
\includegraphics[width=\linewidth]{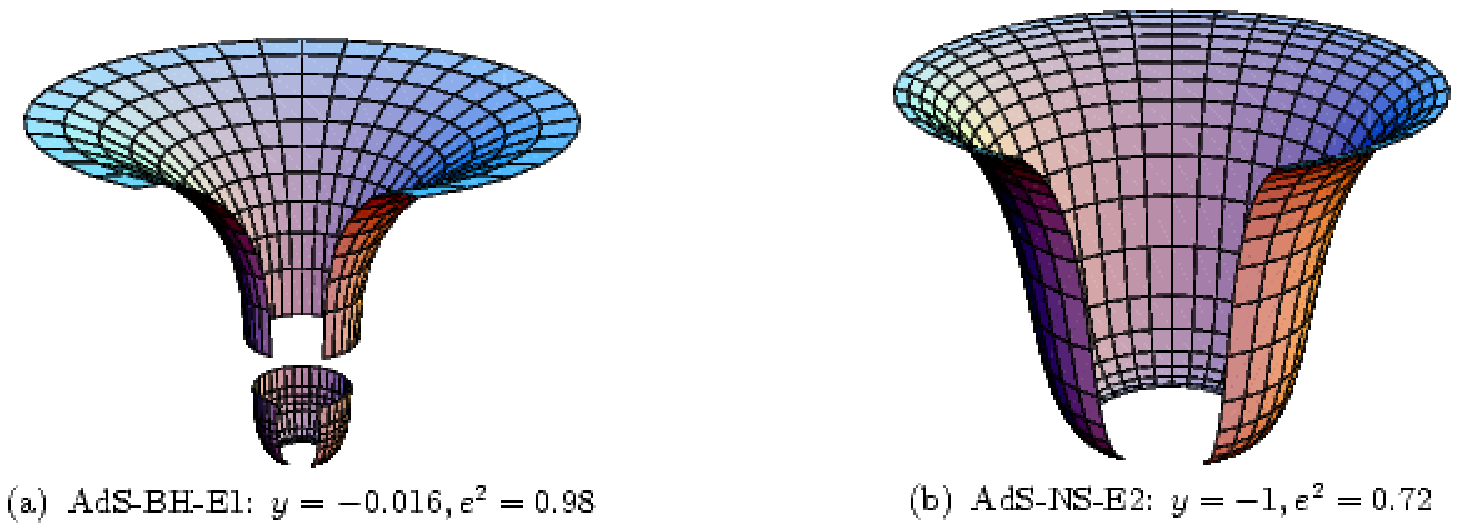}
\caption{Typical embedding
  diagrams of the ordinary space geometry of the asymptotically \adS{}
  spacetimes. There is one type of the diagram for black holes (case (a)),
  and the other one for naked singularities (case (b)). In both cases, the
  embedding diagrams cannot be extended up to infinity. Note that for some
  of the naked-singularity spacetimes (type AdS-NS-1) no region is
  embeddable into the Euclidean space.}
\label{f15}
\end{figure}

\subsection*{A1\enspace\RN\nnd\dS{} black-hole spacetimes}

The embedding diagrams have the same character for all allowed values of
$e$ and $y$; following the method introduced in the discussion of the
effective potential of the geodetical motion, we denote the type as
dS-BH-E1.  A typical embedding is given in Fig.\,\ref{f14}a. The ordinary
space between the outer black-hole horizon $r_{\text{b}+}$ and the
cosmological horizon $r_{\text{c}}$ can always be embedded\Md the embedding
resembles a funnel having a throat near both the horizons. This is the same
behavior as in the case of \Schw\nnd\dS{} black holes (for details see
Ref.\,\cite{Stu-Hle:1999:PHYSR4:}). On the other hand, embeddability of the
space under the black-hole inner horizon is limited from below.

\subsection*{A2\enspace\RN\nnd\dS{} naked-singularity spacetimes}

The embedding diagrams have the same character for all allowed values of
$e$, and $y$; we denote the type as dS-NS-E1, and we show a typical
embedding in Fig.\,\ref{f14}b. The ordinary space is embeddable from a lower
bound given by the function $y_{\text{e(ord)}}(r;e)$ up to the cosmological
horizon.  The embedding diagram has a throat near the cosmological horizon.

\subsection*{B1\enspace\RN\nnd\adS{} black-hole spacetimes}

The embedding diagrams are of the same nature for all black holes; we
denote the type as AdS-BH-E1. A typical embedding diagram is shown in
Fig.\,\ref{f15}a.  The embedding consists of two regions, the inner one
approaches the inner horizon being limited from below. The outer region
approaches the outer horizon being limited from above.

\subsection*{B2\enspace\RN\nnd\adS{} naked-singularity spacetimes}

There are two types of naked singularities with respect to embeddability of
the ordinary space.  The type AdS-NS-E1 has no region that could be
embedded into Euclidean space. The type AdS-NS-E2 has a region that can be
embedded, but it is limited both from below and from above; a typical
embedding is shown in Fig.\,\ref{f15}b.

\section{Optical reference geometry and the inertial forces}\label{opt}

The optical reference geometry enables us to introduce the concept of
gravitational and inertial forces in the framework of general relativity in
a natural way
\cite{Abr:1992:MONNR:,%
      Abr-Nur-Wex:1993:CLAQG:,%
      Abr-Pra:1990:MONNR:,%
      Abr-Nur-Wex:1995:CLAQG:,%
      Agu-etal:1996:CLAQG:RaN,%
      Stu:1990:BULAI:,%
      Abr-Bic:1991:GENRG2:,%
      Son-Abr:1998:JMATP4:,%
      Abr-Mil-Stu:1993:PHYSR4:}.
    In accord with the spirit of general relativity, alternative approaches to
the concept of inertial forces are possible
\cite{Sem:1995:NUOC2:,Jan-Car-Bin:1992:ANNPH1:}, however, the optical
geometry provides a description of relativistic dynamics in accord with
Newtonian intuition.  The optical geometry results from an appropriate
conformal ($3+1$) splitting, reflecting certain hidden properties of the
spacetime under consideration through its geodetical structure. The geodesics
of the optical geometry related to static spacetimes coincide with
trajectories of light, thus being `optically straight'
\cite{Abr-Car-Las:1988:GENRG2:,Abr:1990:MONNR:}. Moreover, the geodesics
are `dynamically straight,' because test particles moving along them are
kept by a velocity-independent force \cite{Abr:1992:MONNR:}, and they are
also `inertially straight,' because gyroscopes carried along them do not
precess along the direction of the motion
\cite{Abr:1992:MONNR:,Abr-Pra:1990:MONNR:}.

The notions of the optical geometry and the related gravitational and
inertial forces are convenient for spacetimes with symmetries, particularly
for stationary (static) and axisymmetric (spherically symmetric) ones.
However, they can be introduced for a general spacetime lacking any
symmetry \cite{Abr-Pra:1990:MONNR:}. Assuming a hypersurface globally
orthogonal to a timelike unit vector field $n^{\kappa}$ and a scalar
function $\Phi$ satisfying the conditions
\bea
  n_{[\kappa} \nabla\!_{\lambda}n_{\mu]} &=& 0,               \label{e60}\\
  n^{\kappa}n_{\kappa} &=& 1,                                 \label{e61}\\
  \dot{a}_{\lambda} &=& n^{\kappa}\nabla\!_{\kappa}n_{\lambda}
    = \nabla\!_{\lambda}                                        \label{e62}
\Phi,
\eea
the \four velocity $u^{\kappa}$ of a test particle of rest mass $m$ can be
uniquely decomposed as
\beq                                                             \label{e63}
  u^{\kappa} = \gamma \left(n^{\kappa} +v\tau^{\kappa} \right),
\eeq
where $\tau^{\kappa}$ is a unit vector orthogonal to $n^{\kappa}$, $v$ is
the speed and $\gamma = (1-v^2)^{-1/2}$ is the Lorentz factor. According to
Abramowicz, Nurowski and Wex~\cite{Abr-Nur-Wex:1993:CLAQG:} so called
ordinary projected geometry (three-space orthogonal to $n^{\kappa}$) can be
introduced by
\beq                                                             \label{e64}
  h_{\kappa \lambda} = g_{\kappa\lambda}+ n_{\kappa}n_{\lambda}
\eeq
and the optical geometry by conformal rescaling
\beq                                                             \label{e65}
  \tilde{h}_{\kappa\lambda} = e^{-2 \Phi}
  \left(g_{\kappa \lambda} +n_{\kappa}n_{\lambda}\right). 
\eeq
Then the projection of the \four acceleration $a^{\bot}_{\kappa} =
h^{\lambda} _\kappa u^{\mu} \nabla\!_{\mu} u_{\lambda}$ can be uniquely
decomposed into terms proportional to zeroth, first and second powers of
$v$, respectively and the velocity change
\beq                                                             \label{e66}
  \dot{v} = \left(e^{\Phi} \gamma v \right)_{,\mu}u^{\mu}. 
\eeq
Thus we arrive at a covariant definition of gravitational and inertial
forces analogous to the Newtonian physics
\beq                                                             \label{e67}
  ma^{\bot}_{\kappa} = G_{\kappa}(v^0) +C_{\kappa}(v^1) +
  Z_{\kappa}(v^2) +E_{\kappa}(\dot{v}),
\eeq
where the first term
\beq                                                             \label{e68}
  G_{\kappa} = - m \nabla\!_{\kappa} \Phi = -m\Phi _{,\kappa},
\eeq
corresponds to the gravitational force, the second term
\beq                                                             \label{e69}
  C_{\kappa} = -m\gamma^2 vn^{\lambda} \left(\nabla\!_{\lambda}
    \tau_{\kappa} - \nabla\!_{\kappa} \tau_{\lambda} \right),
\eeq
corresponds to the Coriolis--Lense--Thirring force, the third term
\beq                                                             \label{e70}
  Z_{\kappa} =
    -m (\gamma v^2)\tilde{\tau}^{\lambda} \tilde{\nabla}\!_{\lambda}
    \tilde{\tau}_{\kappa},
\eeq
corresponds to the centrifugal force, and the last term
\beq                                                             \label{e71}
  E_{\kappa} = -m \dot{v} \tilde{\tau}_{\kappa}
\eeq
corresponds to the Euler force.  Here, $\tilde{\tau}^{\kappa}$ is the unit
vector along $\tau^{\kappa}$ in the optical geometry, and
$\tilde{\nabla}\!_{\kappa}$ is the covariant derivative with respect to the
optical geometry.

In the simple case of static spacetimes with a field of timelike Killing
vectors $\vec{n} \equiv \vec{\xi}_{(t)} = \p/\p t$, we are dealing with
space components that we will denote by Latin indices in the following. The
metric coefficients of the optical reference geometry are given by formula
\beq                                                             \label{e72}
  \tilde{h}_{ik} = \e^{-2\Phi} h_{ik},
\eeq
where $h_{ik}$ are metric components of the three-space ordinary geometry,
and
\beq                                                             \label{e73}
  \e^{2\Phi} = -g_{\mu\nu} \xi^{\mu}_{(t)}\xi^{\nu}_{(t)} = - g_{tt}.
\eeq
In the optical geometry related to a static spacetime, we can define
three-momentum of a test particle
\cite{Abr-Pra:1990:MONNR:,Stu:1990:BULAI:}
\beq                                                             \label{e78}
  \tilde{p}^i = \e^{2\Phi} p^i,
\eeq
and a three-force acting on the particle
\beq                                                             \label{e79}
  \tilde{f}_i = \e^{2\Phi} f_i,
\eeq
where $p^i$, $f_i$ are space components of the \four momentum $p^{\mu}$ and
the \four force $f_\mu$. Then the equation of motion instead of the full
spacetime form
\beq                                                             \label{e77}
  mf_{\mu} = p^{\nu} \nabla\!_{\nu} p_{\mu}.
\eeq
takes the following form in the optical geometry
\beq                                                             \label{e80}
  m\tilde{f}_i = \tilde{p}^j \tilde{\nabla}\!_j \tilde{p}_i
                + \tfrac{1}{2} m^2 \tilde{\nabla}\!_i \Phi,
\eeq
where $\tilde{\nabla}\!_j$ represents the covariant derivative with respect
to the optical geometry. We can see directly that photon trajectories
($m=0$) are geodesics of the optical geometry. The first term on the right
hand side of (\ref{e80}) corresponds to the centrifugal force
\beq                                                             \label{e81}
  \tilde{Z}_k = \tilde{p}^j \tilde{\nabla}\!_j\tilde{p}_k =
                \tilde{p}^j \partial_j \tilde{p}_k -
                \tfrac{1}{2} \tilde{p}^j \tilde{p}^i \partial_k
                \tilde{g}_{ji};
\eeq
the second term corresponds to the gravitational force
\beq                                                             \label{e82}
  -m\tilde{G}_k = \tfrac{1}{2} m^2 \tilde{\nabla}\!_k \Phi =
                  \tfrac{1}{2} m^2 \partial_k \Phi.
\eeq

We shall demonstrate the behavior of the centrifugal force in the
\RN\nnd\aodS{} backgrounds in the simple case of motion of test particles
along circular trajectories; the circular motion will be considered both in
the equatorial plane, and outside of the plane ($r = \text{const}$,
$\theta \neq \pi/2 = \text{const}$). All such orbits are related to the
field of the axial Killing vector $\vec{\xi}_{(\phi)}= \p/\p \phi$. In the
corresponding optical geometry circular orbits can be defined by using the
Killing vector field
\beq                                                             \label{e83}
  \tilde{\xi}_{(\phi)i} = e^{-2\Phi} \xi_{(\phi)i}.
\eeq
Radius of the circular orbit, as measured in the optical geometry,
determines so called radius of gyration that is given by the relation
\cite{Abr-Mil-Stu:1993:PHYSR4:}
\beq                                                             \label{e84}
  \tilde{r}=\left(\tilde{\xi}^i_{(\phi)} \tilde{\xi}_{(\phi)i}\right)^{1/2}
           = r\sin\theta \e^{\Phi},
\eeq
and the unit vector tangent to the orbit is
\beq                                                             \label{e85}
  \tilde{\tau}_i = \tilde{r}^{-1} \tilde{\xi}_{(\phi)i}.
\eeq
The geodetical curvature $\tilde{R}$ of a circle is defined by the relation
\beq                                                             \label{e86}
  \tilde{\tau}^i \tilde{\nabla}\!_i \tilde{\tau}_k =
                -\tilde{R}^{-1} \tilde{\lambda}_k,
\eeq
where the unit vector $\tilde{\lambda}_k$ is the first normal of the
circle. The geodesic can be expressed in the form
\beq                                                             \label{e87}
  \tilde{R}^{-2} = \tilde{r}^{-2} \tilde{h}^{ik}
                   \left(\tilde{\nabla}\!_i \tilde{r} \right)
                   \left(\tilde{\nabla}\!_k \tilde{r} \right).
\eeq
Introducing velocity $v$ of the particle relative to the optical geometry
in the standard Newtonian way
\beq                                                             \label{e88}
  \tilde{p}_k = m v \tilde{\tau}_k,
\eeq
the magnitude of the centrifugal force is given by the relation
\beq                                                             \label{e89}
  Z = \frac{mv^2}{\tilde{R}},
\eeq
that is formally identical with the classical Newtonian relation.

In the static regions of the \RN\nnd\aodS{} spacetimes, the metric
coefficients of the optical reference geometry are given by the relations
\bea
  \tilde{h}_{rr} &=& \e^{-4\Phi},                               \nonumber\\
  \tilde{h}_{\theta\theta} &=& r^2\e^{-2\Phi},                \label{e74}\\
  \tilde{h}_{\phi\phi} &=& r^2 \sin^2\theta\e^{-2\Phi},         \nonumber
\eea
where
\beq                                                             \label{e76}
  \e^{2\Phi} = 1- \frac{2}{r} +\frac{e^2}{r^2} - yr^2.
\eeq
Therefore, the geodetical curvature of the equatorial circular orbits takes
the form
\beq                                                             \label{e90}
  \tilde{R}\left(\theta = \frac{\pi}{2} \right) =
                 r\left(1- \frac{3}{r} +\frac{2e^2}{r^2} \right)^{-1},
\eeq
while for the off-equatorial orbits it reads
\beq                                                             \label{e91}
  \tilde{R}\left(\theta \neq \frac{\pi}{2} \right) =
                r \left[\left(1- \frac{3}{r} +\frac{2e^2}{r^2}\right)^2
                + \left(1- \frac{2}{r} +\frac{e^2}{r^2} -yr^2 \right)
                \cot^2 \theta \right]^{-1/2}.
\eeq
Similarly, the magnitude of the centrifugal force is then given by the
relations
\beq                                                             \label{e92}
  Z\left(\theta= \frac{\pi}{2}\right) =
                 \frac{mv^2}{r} \left(1-\frac{3}{r} +
                 \frac{2e^2}{r^2} \right),
\eeq
and
\beq                                                             \label{e93}
  Z\left(\theta \neq \frac{\pi}{2} \right) =
                \frac{mv^2}{r}\left[\left(
                1- \frac{3}{r} +\frac{2e^2}{r^2}\right)^2 +
                \left(1- \frac{2}{r} +\frac{e^2}{r^2} -yr^2 \right)
                \cot^2 \theta\right]^{1/2}.
\eeq
Notice that the geodetical curvature of circular orbits in the equatorial
plane is independent of the cosmological parameter $y$, while for circular
orbits outside the equatorial plane $y$ enters the expression for
$\tilde{R}$. The same statement holds for the magnitude of the centrifugal
force.

It is important that in the equatorial plane $\tilde{R} \to \infty$ at the
radii corresponding to photon circular orbits. Since $\tilde{R} \to \infty$
there, these circular orbits are geodesics of the optical geometry, and, as
we can directly see from Eq.\,(\ref{e92}), the centrifugal force vanishes
(and changes its sign) at the radii corresponding to photon circular
geodesics.

\section{Embedding of the optical geometry}\label{embed}

Some fundamental properties of the optical geometry can be appropriately
demonstrated by \ed{} of its representative sections
\cite{Kri-Son-Abr:1998:GENRG2:,%
      Stu-Hle:1999:CLAQG:,%
      Stu-Hle:1999:PHYSR4:,%
      Stu-Hle-Jur:2000:CLAQG:,%
      Stu-etal:2001:PHYSR4:}.
Because we are familiar with the Euclidean space, we shall embed the
two-dimensional equatorial plane of the optical geometry associated to the
static regions of the \RN\nnd\aodS{} spacetimes into the three-dimensional
Euclidean space. (Of course, embeddings into other suitably chosen spaces
can also provide interesting information, however, we shall focus our
attention on the most straightforward Euclidean space.)

The line element of the equatorial plane of the optical geometry, given by
the relation
\beq                                                             \label{e94}
  \diff\tilde{\ell}^2_{\text{(RNdS)}} =
    \frac{r^4}{{\left(-yr^4 +r^2 -2r +e^2 \right)}^{2}}\,\diff r^2 +
    \frac{r^4}{\left(-yr^4 +r^2 -2r +e^2 \right)}\,\diff\phi^2,
\eeq
has to be identified with the line element $\diff \ell^2_{\text{(E)}}$,
given by Eq.\,(\ref{e52}). The azimuthal angles can again be directly
identified.  For the radial coordinates, however, we have to put
\beq                                                             \label{e95}
  \rho = r \left(1- \frac{2}{r} +\frac{e^2}{r^2} -yr^2\right)^{-1/2}.
\eeq
It follows immediately from Eq.\,(\ref{e95}) that turning points of the
\ed{} are given by the condition
\beq                                                             \label{e99}
  \oder{\rho}{r} = \frac{r\left(r^2 -3r +2 e^2 \right)}
                {\left(-yr^4 +r^2 -2r +e^2 \right)^{3/2}} = 0.
\eeq
We see directly that the turning points are located just at $r=
r_{\text{ph}+}$ (and $r =r_{\text{ph}-}$ in the case of naked singularities
with parameters $y$ and $e$ chosen appropriately), corresponding to the
radii of photon circular orbits. At $r=r_{\text{ph}+}$, there is a throat
of the embedding diagram, while at $r =r_{\text{ph}-}$ it has a belly. The
radii of the photon circular orbits are important from the dynamical point
of view. At $r>r_{\text{ph}+}$, the dynamics is qualitatively Newtonian
with the centrifugal force directed towards increasing $r$. At
$r=r_{\text{ph}+}$, the centrifugal force vanishes and at $r<
r_{\text{ph}+}$, it is directed towards decreasing $r$. In the field of
naked singularities admitting a circular photon orbit at $r_{\text{ph}-}$,
the centrifugal force vanishes at $r=r_{\text{ph}-}$ and changes its sign
again, i.e., it is directed towards increasing $r$ at $r<r_{\text{ph}-}$.
All of these relations are given by the fact that the `effective potential'
of the photon motion, the Euclidean coordinate $\rho$ of the embedding, and
the centrifugal force, all of them are determined by the azimuthal metric
coefficient of the optical geometry $\tilde{h}_{\phi\phi}$.

The embedding diagrams can be effectively constructed using a parametric
form of the embedding formula $z(\rho) = z(\rho(r))$, with $r$ being the
parameter. Since
\beq                                                             \label{e96}
  \oder{z}{\rho} = \oder{z}{r} \oder{r}{\rho},
\eeq
we obtain 
\beq                                                             \label{e97}
  \left(\oder{z}{r} \right)^2 = \left(1- \frac{2}{r}
                +\frac{e^2}{r^2} - yr^2 \right)^{-2}
                - \left(\oder{\rho}{r} \right)^2,
\eeq
and finally we arrive at the embedding formula
\beq                                                             \label{e98}
  \oder{z}{r} = \pm \frac{r \left[-yr^6 +4r^3 -(3e^2 +9)r^2
                + 12 e^2r -4e^4 \right]^{1/2}}
                {\left(-yr^4 +r^2 -2r +e^2 \right)^{3/2}}.
\eeq

The condition of embeddability
\beq                                                            \label{e100}
  4r^3 -(3 e^2 +9)r^2 +12 e^2 r -4 e^4 -yr^6 \geq 0
\eeq
can be expressed in the form
\beq                                                            \label{e101}
  y\leq y_{\text{e(opt)}}(r;e)
    \equiv\frac{4r^3 -(3 e^2 +9)r^2 +12 e^2 r -4 e^4}{r^6},
\eeq
where the function $y_{\text{e(opt)}}(r;e)$ determines the limits of
embeddability, i.e., the boundaries of \ed{} if they do not coincide with
the horizons of the spacetime under consideration. The asymptotic behavior
of $y_{\text{e(opt)}}(r;e)$ is given by the relations
\beq
  y_{\text{e(opt)}}(r \to 0,e) \to - \infty, \quad
  y_{\text{e(opt)}}(r \to \infty, e) \to 0.
\eeq
The zero points of the function $y_{\text{e(opt)}}(r;e)$ are given by the
relation
\beq                                                             \label{102}
  e^2 = e^2_{\text{z(e(opt))}\pm}(r) \equiv
        \frac{3r(4 -r) \pm r\sqrt{r(9r -8)}}{8},
\eeq
the function $e^2_{\text{z(e(opt))}\pm}(r)$ is drawn in Fig.\,\ref{f11}.
For $e=0$, the function $y_{\text{e(opt)}}(r;e=0)$ has a zero point at
$r_{\text{(opt)}} = 9/4$. Since
\beq                                                            \label{e103}
  \pder{y_{\text{e(opt)}}(r,e)}{r} = \frac{-12(r -e^2)(r^2 -3r +2e^2)}{r^7},
\eeq
we can see that the local extrema of $y_{\text{e(opt)}}(r;e)$ are located
at
\beq                                                     \label{e104}
  r_1 = e^2, \quad r_2 = r_{\text{ph}-}(e), \quad r_3 = r_{\text{ph}+}(e),
\eeq
where $r_{\text{ph}\pm}(e)$ determine radii of photon circular orbits. Note
that both $r_2(e)$ and $r_3(e)$ are implicitly given by Eq.\,(\ref{e68}).
The functions $e^2_{\text{z(h)}}(r)$, and $e^2_{\text{e(opt)}}(r) = r$ are
drawn in Fig.\,\ref{f11}.  Considering $\p^2y_{\text{e(opt)}}(r;e)/\p r^2$,
we can summarize the character of the local extrema at $r = r_{1,2,3}$ in
the following way. For $0< e^2 <1$ there is a local minimum at $r_1$, and
local maxima at $r_2, r_3$. For $e^2 =1$, $r_1$ and $r_2$ coalesce at an
inflex point. For $1< e^2 < 9/8$, there are local maxima at $r_1$ and
$r_3$, and a local minimum at $r_2$. For $e^2 = 9/8$, $r_2$ and $r_3$
coalesce at an inflex point. For $e^2 > 9/8$ there is only a local maximum
at $r_1$.  The properties of the \ed{} are determined by the behavior of
$y_{\text{e(opt)}}(r,e)$ and $y_{\text{h}}(r;e)$. The embedding is
possible, if
\beq                                                            \label{e105}
  y \leq y_{\text{e(opt)}}(r;e) \leq y_{\text{h}}(r;e).
\eeq
In order to obtain a classification of the \RN\nnd\aodS{} spacetimes
according to the properties of the embedding diagrams of their optical
geometry, and to determine distribution of different types of these
spacetimes in the parameter space $e^2$-$y$, we have to introduce an
additional relevant quantity, namely the value of the local extreme of
$y_{\text{e(opt)}}(r;e)$ at $r_1 = e^2$.  This extreme is given by
\beq                                                            \label{e106}
  y_{\text{e(opt)(extr)}}(e) = \frac{e^2 -1}{e^8}
\eeq
and must be compared with $y_{\text{h(min)}}(e)$ and $y_{\text{h(max)}}(e)$
at $ r=r_2 =r_{\text{ph}-}$, and $r =r_3 =r_{\text{ph}+}$.

Properties of the embedding diagrams can be summarized using behavior of
the functions $y_{\text{e(opt)}}(r;e)$ and $y_{\text{h}}(r;e)$ in three
qualitative different cases according to the charge parameter $e$: (a)~$0 <
e^2 \leq 1$, (b)~$1 < e^2 \leq 9/8$, (c)~$e^2 > 9/8$. Behavior of these
functions is illustrated in Fig.\,\ref{f17}. Analysis of these
characteristic functions shows that there are nine types of the
\RN\nnd\aodS{} spacetimes with qualitatively different behavior of the
embedding diagrams of the optical geometry. We shall define the spacetimes
in the following way:

\begin{description}
\item[dS-BH-O1] The embedding is continuously extended between two
  boundaries given by the function $y_{\text{e(opt)}}(r;e)$ located between
  the outer black-hole and cosmological horizons; it has a throat.
\item[dS-BH-O2] The embedding has two parts. The inner one is located
  under the inner black-hole horizon, and has no turning point. The outer
  one is extended between the two boundaries given by the function
  $y_{\text{e(opt)}}(r;e)$, and it is located between the outer black-hole
  horizon and the cosmological horizon. The outer part has a throat.
\item[dS-NS-O1] No embedding is possible.
\item[dS-NS-O2] The embedding is continuously extended between two
  boundaries given by the function $y_{\text{e(opt)}}(r;e)$, and has a
  belly and a throat.
\item[dS-NS-O3] The embedding is continuously extended between two
  boundaries given by the function $y_{\text{e(opt)}}(r;e)$. It has no
  turning point.
\item[AdS-BH-O1] The embedding is continuously extended above the outer
  horizon between the boundary given by the function $y_{\text{e(opt)}}$
  and infinity, having a throat.
\item[AdS-NS-O1] The embedding is continuous, located between the inner
  boundary given by the function $y_{\text{e(opt)}}(r;e)$ and infinity. It
  has a belly and a throat.
\item[AdS-NS-O2] The embedding has two parts. The inner one has two
  boundaries given by the function $y_{\text{e(opt)}}(r;e)$ and a belly,
  the outer one is located between the boundary given by the function
  $y_{\text{e(opt)}}(r;e)$, and infinity, and has a throat.
\item[AdS-NS-O3] The embedding extends between the inner boundary given by
  $y_{\text{e(opt)}}(r;e)$ and infinity; it has no turning point.
\end{description}

\begin{figure}[p]
\centering
\includegraphics[width=.56\linewidth]{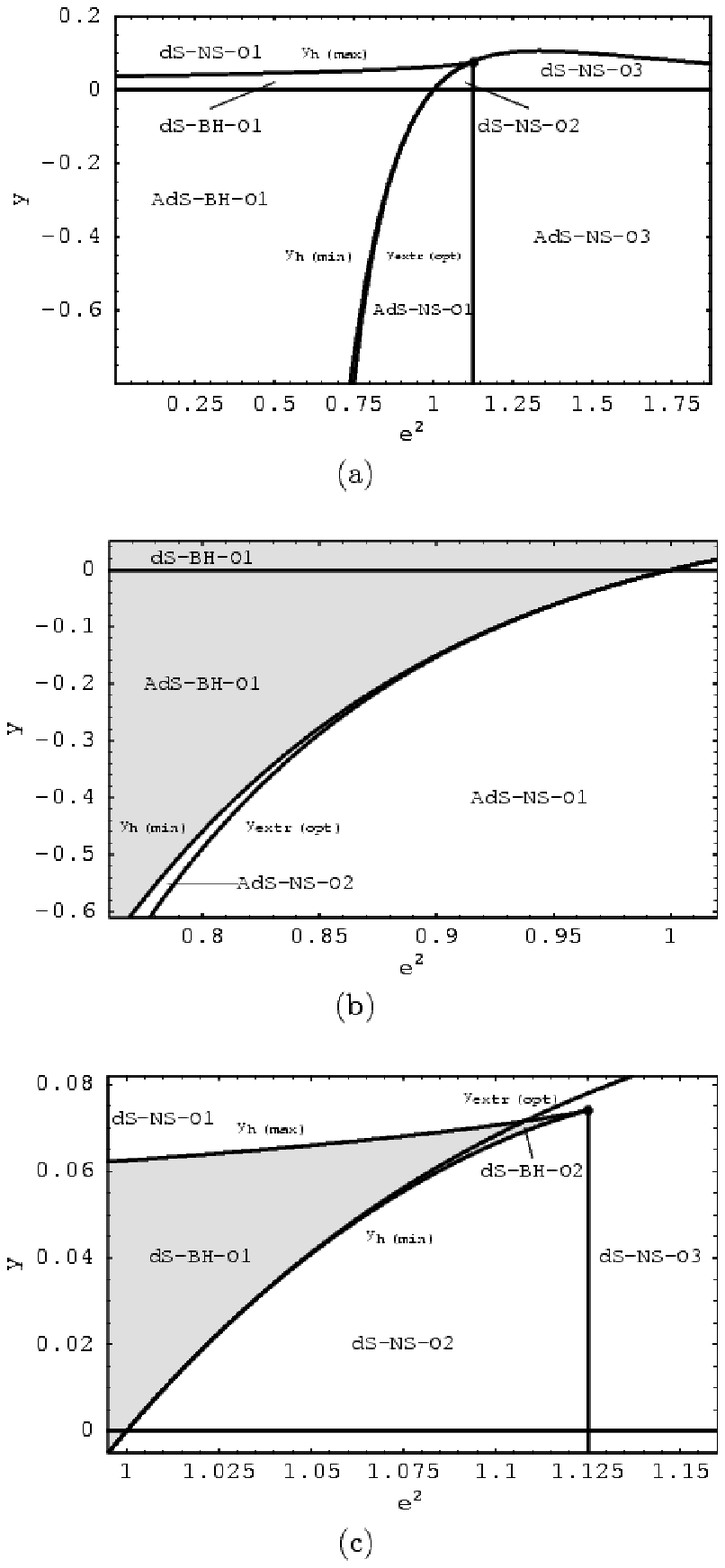}
\caption{The classification of the \RN\nnd\aodS{} spacetimes according
  to the properties of the embedding diagrams of the associated optical
  reference geometry. The functions $y_{\text{h(min)}}(e)$ and
  $y_{\text{h(max)}}(e)$ limit the region of black-hole spacetimes (shaded)
  in the parameter space $e^2$-$y$, and they separate spacetimes of
  different type of the embeddings, along with the function
  $y_{\text{extr(opt)}}(e)$, and the lines $y = 0$ (for $e^2 \geq 0$) and
  $e^2 = 9/8$ (for $y < 2/27$).}
\label{f16}
\end{figure}

\begin{figure}[p]
\centering
\includegraphics[width=.57\linewidth]{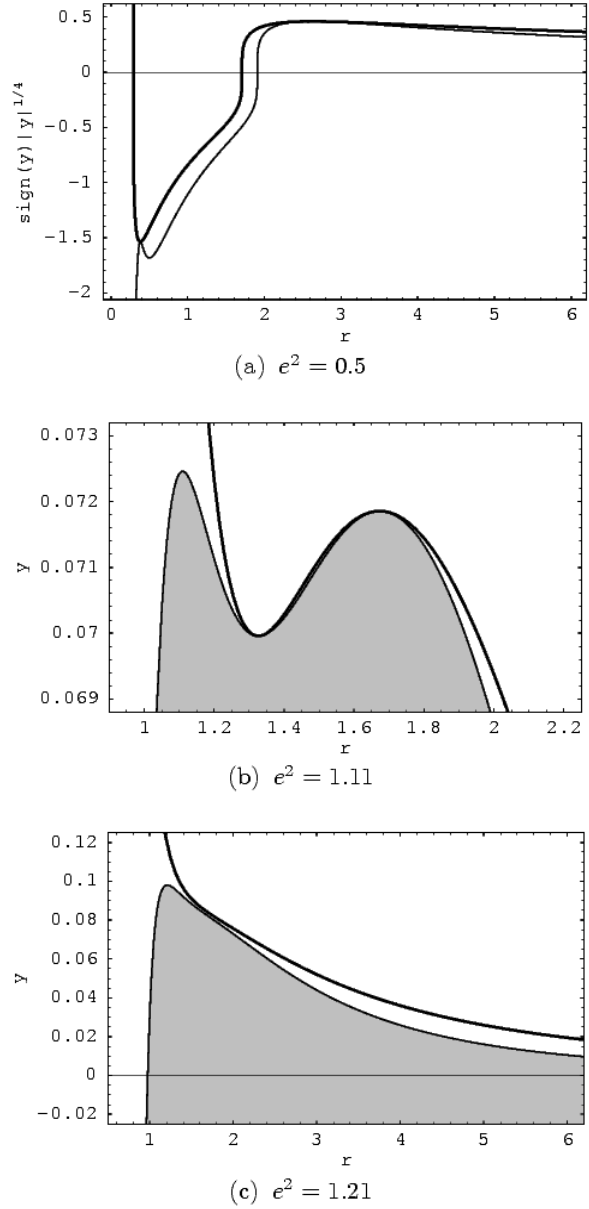}
\caption{Embeddability of
  the optical reference geometry associated with the \RN\nnd\aodS{}
  spacetimes. According to the value of the charge parameter $e^2$, there
  are three qualitatively different types of the behaviour of the functions
  $y_{\text{h}}(r;e)$ (bold lines) and $y_{\text{e(opt)}}(r;e)$ (thin solid
  lines) determining the regions of embeddability. The regions of
  embeddability are shaded. Turning points of the embedding diagrams
  correspond to the photon circular orbits located at radii given by the
  local extrema of $y_{\text{h}}(r;e)$.}
\label{f17}
\end{figure}

\begin{figure}[t]
\centering
\includegraphics[width=.85\linewidth]{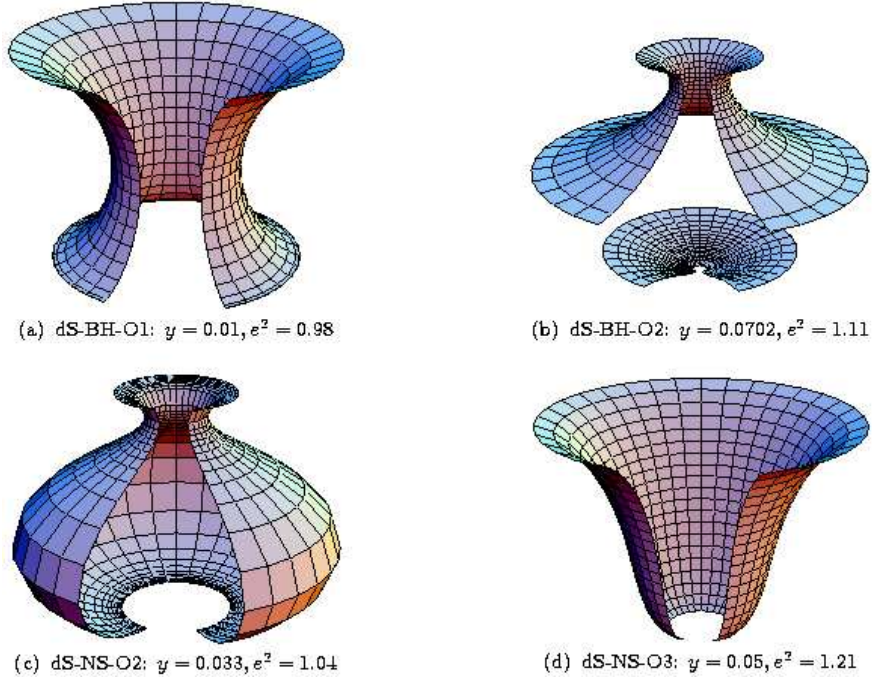}
\caption{Typical embedding diagrams of the optical reference geometry
  of the asymptotically \dS{} spacetimes. In the black-hole spacetimes a
  limited region between the outer black-hole horizon and the cosmological
  horizon can be embedded, and some part of the region between the
  singularity and the inner black-hole horizon can be embedded in the type
  of dS-BH-O2 (case (b)). For naked-singularity spacetimes, a limited
  region between the singularity and the cosmological horizon can be
  embedded. However, for the type of dS-NS-O1, no region of the optical
  geometry can be embedded into the Euclidean space. Turning points of the
  embedding diagrams correspond to the radii of the photon circular
  geodesics, which are independent of the cosmological parameter $y$. The
  centrifugal force vanishes there; within the upward (inward) sloping
  areas of the embedding diagram the centrifugal force is outward (inward)
  directed.}
\label{f18}
\end{figure}

\begin{figure}[t]
\centering
\includegraphics[width=.85\linewidth]{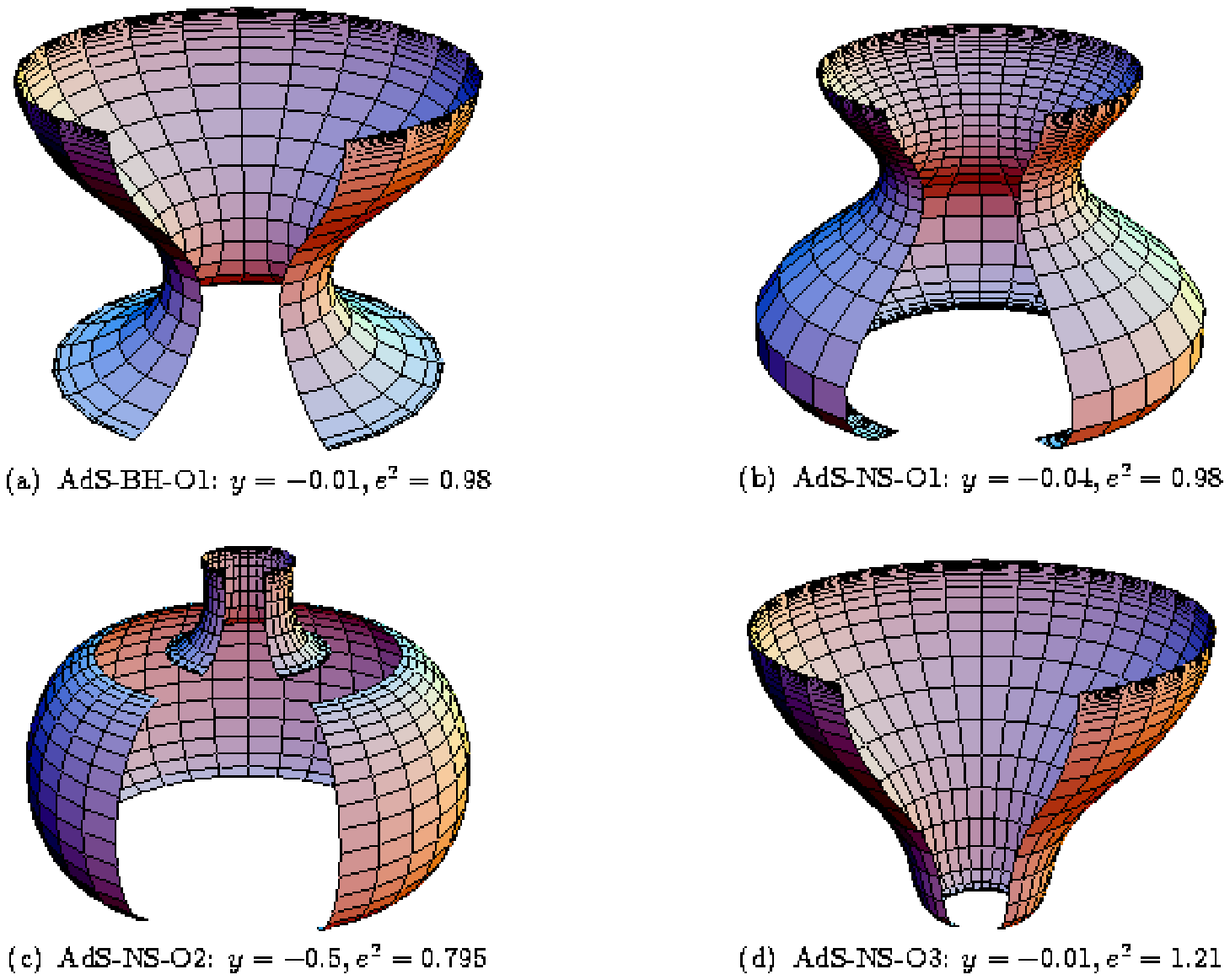}
\caption{Typical embedding
  diagrams of the optical reference geometry of the asymptotically \adS{}
  spacetimes. There is one type of the embedding for black holes (case
  (a)), and three types of the embedding for naked singularities (cases (b)
  through (d)).  In all of these cases, the embedding is possible for $r
  \to \infty$, however, it is shrunk to a finite value of the embedding
  coordinate $\rho$, depending on the cosmological parameter $y$. Turning
  points of the embedding diagrams correspond to the radii of the photon
  circular geodesics, which are independent of $y$. Behavior of the
  centrifugal force is governed by the same rules as in the case of
  asymptotically \dS{} spacetimes. }
\label{f19}
\end{figure}

In the parameter space $e^2$-$y$, distribution of types of the
\RN\nnd\aodS{} spacetimes with different properties of the \ed{} of their
optical geometry is given in Fig.\,\ref{f16}. Typical \ed{} are constructed
by a numerical code and presented in Figs~\ref{f18} and \ref{f19}.
Properties of the embedding diagrams of the optical geometry can be
summarized in the following way.

\subsection*{A1\enspace\RN\nnd\dS{} black-hole spacetimes}

There are two types of the black-hole spacetimes (dS-BH-O1 and 2). For both
of them, the embedding diagram has a throat between the outer black-hole
horizon and the cosmological horizon.  None of the embedding diagrams can
reach any of the event horizons. For the spacetimes of the type dS-BH-O1
(with $e^2 \leq 1$), the embedding is not possible under the inner horizon,
while in the spacetimes dS-BH-O2 ($1 < e^2 \leq 9/8$), some part of the
region between the singularity at $r = 0$ and the inner horizon can be
embedded. The character of the embedding in the case of dS-BH-O1 spacetimes
is just the same as in the case of \Schw\nnd\dS{} black holes. For dS-BH-O2
spacetimes the difference is given by the inner part of the embedding. In
the case of Reissner-Nordstr\"om black-hole spacetimes, the embedding can
be extended up to $r \to \infty$. Typical embedding diagrams are
constructed by a numerical procedure and given in Fig.\,\ref{f18}a
(dS-BH-O1) and Fig.\,\ref{f18}b (dS-BH-O2).

\subsection*{A2\enspace\RN\nnd\dS{} naked-singularity spacetimes}

There are three types of naked-singularity spacetimes. In the type dS-NS-O1
spacetimes no embedding is admitted. In the type dS-NS-O2 spacetimes, the
embedding has a belly and a throat (Fig.\,\ref{f18}c). In the type dS-NS-O3
spacetimes, there are no turning points of the embedding diagrams
(Fig.\,\ref{f18}d). The embedding does not reach neither the singularity
and the cosmological horizon in both dS-NS-2 and 3 spacetimes. This is the
only difference with the embeddings of \RN{} naked-singularity spacetimes,
where the embedding can be extended up to $r \to \infty$.

\subsection*{B1\enspace\RN\nnd\adS{} black-hole spacetimes}

There is only one type of the black-hole spacetimes denoted AdS-BH-O1.  A
typical embedding is shown in Fig.\,\ref{f19}a. The embedding extends from
some radius above the outer horizon up to infinity and has a throat. The
embedding is not allowed in the region under the inner horizon.  Notice
that the \ed{} of both black-hole and naked-singularity backgrounds in the
asymptotically \adS{} universe has a specific property given by the
definition of the embedding coordinate $\rho$ (see Eq.\,(\ref{e95})).  The
\ed{} cover whole the asymptotic part of spacetime in a restricted part of
the Euclidean space. There is
\beq                                                            \label{e107}
  \rho(r \to \infty) = (-y)^{-1/2}.
\eeq

Clearly, with decreasing attractive \cc{} the embedding diagram is deformed
with increasing intensity. The circles of $r = \text{const}$ are
concentrated with an increasing density around $\rho = (-y)^{-1/2}$ as $r
\to \infty$.  This behavior is exactly the same as in the case of
\Schw\nnd\adS{} black holes, and it represents the only difference with
respect to the case of \RN{} black holes.

\subsection*{B2\enspace\RN\nnd\adS{} naked-singularity spacetimes}

There are three types of naked-singularity spacetimes. In the type
AdS-NS-O1 spacetimes, the embedding is continuously extended between the
inner boundary given by the function $y_{\text{e(opt)}}(r;e)$ and infinity,
and has a belly and a throat (Fig.\,\ref{f19}b). In the type of AdS-NS-O2
spacetimes, the embedding is separated into two parts\Md the inner one has
a belly, the outer one has a throat (Fig.\,\ref{f19}c). In the type of
AdS-NS-O3 spacetimes, the embedding is continuously extended between the
inner boundary given by $y_{\text{e(opt)}}(r;e)$ and infinity, but it has
no turning point (Fig.\,\ref{f19}d). The difference with respect to
embeddings of the \RN{} naked-singularity spacetimes is given by AdS-NS-O2
spacetimes, where the embedding has two parts, and by the asymptotic
behavior of the embedding diagram in all the type AdS-NS-O1 through 3
spacetimes.

\section{Asymptotic behavior of the optical geometry}\label{asympt}

The reason why the embedding into the Euclidean space is not possible in
some parts of the optical (or ordinary) geometry is explained in
\cite{Kri-Son-Abr:1998:GENRG2:,Stu-Hle:1999:PHYSR4:}.  However, the optical
geometry is still well defined outside the regions of the embeddability
into Euclidean space nearby the event horizons. It is possible to
demonstrate its properties by the behavior of the proper lengths along the
radial direction. In the optical geometry, the proper radial length
coincides with the well known Regge\nnd Wheeler `tortoise' coordinate
\beq                                                            \label{e108}
  r^{\ast} = \int\left(1- \frac{2}{r} +\frac{e^2}{r^2}
             -yr^2\right)^{-1} \diff r.
\eeq
This gives direct relevance of the `tortoise' coordinate in the optical
space, since it can be shown that in the black-hole spacetimes, the outer
black-hole horizon and the cosmological horizon are infinitely far away in
the optical geometry. At $r \sim r_{\text{b}+}$, there is $r^{\ast} \sim
+\ln|r-r_{\text{b}+}| \to - \infty$, while at $r \sim r_{\text{c}}$, there
is $r^{\ast} \sim - \ln|r_{\text{c}} - r| \to + \infty$. On the other hand,
in the ordinary geometry, the horizons are located at a finite proper
radial distance
\beq                                                             \label{109}
  \tilde{r} = \int\left(1- \frac{2}{r} +\frac{e^2}{r^2} -yr^2
              \right)^{-1/2} \diff r;
\eeq
at $r \sim r_{\text{b}+}$, $\tilde{r} \sim \sqrt{r-r_{\text{b}+}}$, and at
$r \sim r_{\text{c}}$, $\tilde{r} \sim \sqrt{|r_{\text{c}} -r|}$. In the
\RN\nnd\dS{} spacetimes, the optical geometry extends infinitely beyond the
limit of embeddability, approaching asymptotically the geometry
\beq                                                            \label{e110}
  \diff\tilde{\sigma}^2 \approx \diff r^{\ast^2} + C_1 \exp
                \left[-r^{\ast}/C(r_{\text{b}+}) \right]
                \left(\diff \theta^2 +\sin^2 \theta\,\diff \phi^2 \right)
\eeq
for $r \to r_{\text{b}+}$, $r^{\ast} \to - \infty$, and
\beq                                                     \label{e111}
  \diff \tilde{\sigma}^2 \approx \diff r^{\ast^2} + C_2 \exp
        \left[-r^{\ast}/C(r_{\text{c}})\right] \left(\diff \theta^2
        + \sin^2 \theta\,\diff\phi^2 \right)
\eeq
for $r \to r_{\text{c}}$, $r^{\ast} \to + \infty$; $C_1$, $C_2$ and
$C(r_{\text{b}+})$, $C(r_{\text{c}})$ are constants given in terms of the
parameters of the spacetime.

For \RN\nnd\adS{} black holes, the optical space has again the property
that at $r \sim r_{\text{b}+}$, there is $r^{\ast} \sim \ln|r
-r_{\text{b}+}| \to - \infty$, while for $r \to \infty$, there is $r^{\ast}
\sim (-y)^{-1} \ln r \to + \infty$, and the asymptotical behavior of the
optical geometry is determined by formulae similar to Eqs~(\ref{e110}) and
(\ref{e111}), respectively.

In the vicinity of the inner black-hole horizon, $r \to r_{\text{b}-}$,
there is $r^{\ast} \sim - \ln|r_{\text{b}-} -r|$ and the optical geometry
has the asymptotical form similar to (\ref{e110}). On the other hand, for
$r \to 0$, there is $r^{\ast} \sim r^3/3e^2 \to 0$ independently of
the value of $y \neq 0$. At $r \to 0$, the asymptotic form of the optical
geometry of the spacetimes with $e^2 \neq 0$ is given by
\beq                                                            \label{e112}
  \diff\tilde{\sigma}^2 \approx \diff r^{\ast^2} + \frac{r^4}{e^2}
              \left(\diff \theta^2 +\sin^2 \theta\,\diff\phi^2 \right).
\eeq
In the naked singularity spacetimes, the asymptotic form of $r^{\ast}$ and
$\diff \tilde{\sigma}^2$ at $r \to r_{\text{c}}$ (for $y > 0$), and at $r
\to \infty$ (for $y < 0$), respectively, is the same as in the black-hole
spacetimes, i.e., $r^{\ast} \to + \infty$ and there is no other divergence
of the `tortoise' coordinate in the naked-singularity spacetimes. There is
$r^{\ast} \sim r^3/3e^2 \to 0$ for $r \to 0$. At $r \sim 0$, the optical
geometry is again determined by the formula (\ref{e112}).

\section{Concluding remarks}\label{conrem}

The \RN\nnd\aodS{} spacetimes can be separated into eleven types of
spacetimes with qualitatively different character of the geodetical motion.
Properties of the motion can be summarized and compared with the properties
of the motion in the \Schw\nnd\dS{} and the \RN{} spacetimes in the
following way.

(1)~The motion above the outer horizon of black-hole backgrounds has the
same character as in the \Schw\nnd\aodS{} spacetimes for both
asymptotically \dS{} ($y>0$) and \adS{} ($y<0$) spacetimes.  Namely, there
is only one static radius (giving limits on existence of circular
geodetical motion) for $y>0$, and no static radius for $y<0$. No static
radius is possible under the inner black-hole horizon for both $y>0$ and
$y<0$, no circular geodesics are possible there.  Only one photon circular
orbit exists above the outer horizon for both $y>0$, and $y<0$; its radius
is, moreover, independent of $y$. No photon circular orbit can exist under
the inner black-hole horizon for both $y>0$, and $y<0$.  From the
astrophysical point of view, the most important are stable circular orbits
as they allow accretion processes in the disk regime. In all of the
asymptotically \adS{} black-hole spacetimes, the stable circular orbits
exist in a region above the outer horizon and can be extended up to $r \to
\infty$ in the limit of ultrarelativistic particles. In the case of
asymptotically \dS{} black-hole spacetimes, there is a region of stable
circular orbits limited both from below and from above by the function
$y_{\text{ms}}(r;e)$. The stable circular orbits can exist only for small
values of the cosmological parameter $y < y_{\text{ms(bh)}} = 0.000693$,
with allowed values of the charge parameter $e^2 < e^2_{\text{ms(bh)}} =
1.000695$. The presence of an outer marginally stable circular geodesic
allows outflow of matter from accretion disks and is, therefore, of high
astrophysical importance \cite{Stu-Sla-Hle:2000:ASTRA:}.

(2)~The motion in the naked-singularity backgrounds has similar character
as the motion in the field of \RN{} naked singularities.  However, in the
case of $y>0$, two static radii can exist, while the \RN{} naked
singularities contain one static radius only.  The outer static radius
appears due to the effect of the repulsive cosmological constant. On the
other hand, the inner static radius, located nearby the ring singularity,
survives even in the presence of an attractive \cc. Two photon circular
orbits exist in naked singularity backgrounds for both $y>0$, and $y<0$, if
$1 < e^2 < 9/8$, just as in the field of \RN{} naked singularities.  Stable
circular orbits exist in all of the naked-singularity spacetimes.  In the
spacetimes with $e^2 > 5/4$ and $y < 0$, there are stable circular
geodesics only. On the other hand, if $e^2 < 5/4$ and $y_{\text{ms(min)}} <
y < y_{\text{ms(max)}}$, there are even two separated regions of stable
circular geodesics, with the inner one being limited by the inner static
radius from bellow, where particles with zero angular momentum (in stable
equilibrium positions) are located. In the asymptotically \dS{}
naked-singularity spacetimes, two regions of stable circular orbits can
exist, if $e^2 < 275/216$, and $y < 0.00174$; otherwise, the inner region
of stable circular orbits survives. More details can be extracted directly
from Fig.\,\ref{f3}.

(3)~In the black-hole spacetimes, escape photon cones has the same
character as in the \Schw\nnd\aodS{} spacetimes. In the asymptotically
\adS{} spacetimes, the capture cone remains nonzero as $r \to \infty$ due
to the effect of the attractive cosmological constant.  In the
naked-singularity spacetimes, the escape photon cones are determined by the
presence of the unstable photon circular orbits\Md in the region located
under the radius of the unstable photon circular orbit, directional angles
corresponding to bound photons appear. The cone of bound photons is most
extended at the radius corresponding to the stable circular photon orbit
(note that photon circular orbits exist, if $e^2 < 9/8$). In spacetimes
with $e^2 > 9/8$, all photons escape, except those radially incoming into
the singularity at $r=0$.

The \ed{} of both ordinary and optical reference geometry give clear
illustration of the influence of both the charge and cosmological parameter
on the structure of the \RN\nnd\aodS{} spacetimes.

(1)~For the ordinary geometry, the embedding is impossible for
\RN\nnd\adS{} naked singularities with $y< y_{\text{e(ord)(min)}}(e)$.  For
all values of $y<0$, and $y>0$, the embedding is impossible in vicinity of
$r=0$. In the asymptotically \adS{} black-hole and naked-singularity
spacetimes, the embeddability is limited from above, too. In the
asymptotically \dS{} black-hole and naked-singularity spacetimes, the
embeddability is possible up to the cosmological horizon. The embedding
diagram resembles a funnel with a throat near the cosmological horizon\Md
with $y$ increasing, the funnel becomes shrunk and flattened.

(2)~For the optical reference geometry, the embedding is possible for all
of the \RN\nnd\adS{} black holes and naked singularities from the below
limit nearby $r=0$ up to infinity, although, it is deformed by shrinking of
the asymptotic region to vicinity of finite value of the embedding radial
coordinate $\rho = (-y)^{-1/2}$. On the other hand, the embeddability is
not possible for the \RN\nnd\dS{} naked singularity spacetimes with $e^2 >
9/8$ and $y > (e^2 -1)/e^8$. Further, for the naked singularity spacetimes
with $e^2 < 9/8$, and both $y>0$ and $y<0$, the \ed{} have a throat and a
belly. In some cases the embedding for naked singularities with $y<0$ is
discontinuous, it consists from two parts.  For black holes (with both $y >
0$ and $y < 0$), the embedding diagram has a throat, the embedding cannot
reach both the inner and outer black-hole horizon. If $y>0$, the embedding
cannot reach the cosmological horizon, too. If $y<0$, no part of the space
under the inner horizon can be embedded.

Embedding diagrams of the optical geometry give an important tool of
visualization and clarification of the dynamical behavior of test
particles moving along equatorial circular orbits: we imagine that the
motion is constrained to the surface
$z(\rho)$~\cite{Kri-Son-Abr:1998:GENRG2:}.  The shape of the embedding
surface $z(\rho)$ is directly related to the centrifugal acceleration.
Within the upward sloping areas of the embedding diagram, the centrifugal
acceleration points towards increasing values of $r$, and the dynamics of
test particles has an essentially Newtonian character. However, within the
downward sloping areas of the embedding diagrams, the centrifugal
acceleration has a radically non-Newtonian character as it points towards
decreasing values of $r$.  Such a kind of behavior appears where the
diagrams have a throat and a belly.  At the turning points of the embedding
diagrams, where $\diff z/\diff\rho = 0$, the centrifugal acceleration
vanishes and changes its sign.

We can understand this connection between the centrifugal force and the
embedding of the optical space in terms of the radius of gyration
representing rotational properties of rigid bodies
\cite{Abr-Mil-Stu:1993:PHYSR4:}.  In Newtonian physics, the gyration radius
$\tilde{r}$ is defined by the relation
\beq
  \tilde{r} = \sqrt{\frac{\ell}{\Omega}}.                    \label{rogsam}
\eeq
where $\ell = J/M$ is the specific angular momentum of a rigid body ($J$ is
angular momentum of the body, $M$ is its mass) rotating with angular
velocity $\Omega$, i.e., it is defined as the radius $\tilde{r}$ of the
circular orbit on which a point-like particle having the same mass $M$ and
angular velocity $\Omega$ would have the same angular momentum
$J=M{\tilde{r}}^2\Omega$.

In Newtonian physics, radius of gyration equals both the circumferential
radius and the radius given by proper radial distance from the rotational
axis, however, in General Relativity they differ. The radius of gyration is
convenient for understanding the dynamical effects of rotation in the
framework of General Relativity, as the direction of increasing $\tilde{r}$
defines the \emph{local outward direction} of these effects.  The surfaces
$\tilde{r} = \text{const}$, called von~Zeipel cylinders, proved to be a
very useful concept in the theory of rotating fluids in stationary, axially
symmetric spacetimes. In Newtonian physics, these are ordinary straight
cylinders but their shape is deformed by general-relativistic effects and
their topology may be noncylindrical.  There is a critical family of
self-crossing von Zeipel surfaces~\cite{Koz-Jar-Abr:1978:ASTRA:}.

The close relation of the centrifugal force, and the embedding diagrams of
the equatorial plane of the optical geometry follows directly from the fact
that the embedding diagrams are expressed in terms of the radius of
gyration
\beq
  [\tilde{h}_{\phi\phi}(\theta=\pi/2)]^{1/2} = \rho = \tilde{r},
\eeq
and the radial component of the centrifugal force in the equatorial plane
is also related to the radius of gyration
\beq
  Z_{\text{R}}(r) \sim \tilde{r}^{-1}\partial_r \tilde{r}.
\eeq

The turning points of the embedding diagrams determine both radii where the
centrifugal force changes its sign and the radii of cusps where the
critical von Zeipel surfaces are self-crossing. Therefore, the embedding
diagrams also reflect the properties of perfect fluid orbiting black holes
or naked singularities.

We can conclude that in most cases the phenomena connected with geodetical
motion and \ed{} in the \RN{} and \Schw\nnd\aodS{} spacetimes are
incorporated into the corresponding phenomena in the \RN\nnd\aodS{}
spacetimes in an additive way.  However, in some cases, new phenomena
appear as a qualitatively new result of the interplay of the effect of
appropriately tuned values of the electric charge and the \cc.

For the geodetical motion, the additive way is realized in most of the cases
considered here. The qualitatively new features caused by an interplay of
the charge and the cosmological constant are the following: the existence
of asymptotically \dS{} black holes with unstable circular geodesics only,
the existence of asymptotically \dS{} naked singularities with an internal
region of stable circular geodesics and an external region of unstable
circular geodesics (if $e^2 < 9/8$, these regions are separated, if $e^2 >
9/8$, they are continuously matched).

For the embedding diagrams, among these qualitatively new phenomena the
following can be ascribed: the nonexistence of \ed{} of the ordinary space
for some naked singularities with $y<0$, and nonexistence of \ed{} of the
optical space for some naked singularities with $y>0$. Existence of
separated parts of the \ed{} of the optical space of some naked
singularities with $y<0$.

In the presented work, attention has been focused on embedding of the
ordinary and optical geometry and the geodetical motion in the
\RN\nnd\aodS{} spacetimes, reflecting the combined effect of an electric
charge and a nonzero \cc{} an the character of black-hole and
naked-singularity spacetimes.  A simple case of equilibrium positions of
charged (and spinning) test particles in these spacetimes has been
discussed in \cite{Stu-Hle:2001:PHYSR4:}, where it has been shown that the
equilibria are independent of the spin of the test particles. The motion of
charged test particles is a more complex problem, and it is under study at
present.

The combined effect of rotation and a nonzero \cc{} in the black-hole and
naked-singularity spacetimes has been extensively studied for the
equatorial photon motion \cite{Stu-etal:1998:PHYSR4:,Stu-Hle:2000:CLAQG:}.
The general geodetical motion and embeddings of the ordinary and optical
geometry are under study at present.

\begin{ack}
  This work was partly supported by the GA\v{C}R grant No.~202/02/0735/A,
  the Committee for collaboration of Czech Republic with CERN and the
  Bergen Computational Physics Laboratory in the framework of the European
  Community\Md Access to Research Infrastructure action of the Improving
  Human Potential Programme. The authors would like to express their
  gratitude to the Theory Division of CERN, where an essential part of the
  work has been done, and to the BCPL at the University of Bergen, where
  the work has been finished, for perfect hospitality.
\end{ack}


\end{document}